\documentclass[
               DIV=15,
               ]{scrartcl}  
\usepackage{color} 
\usepackage{psfrag} 
\usepackage{subfig}
\usepackage{hyperref}
\usepackage{amsmath}
\usepackage{amssymb}
\usepackage{stmaryrd}

\usepackage{xfrac}
\usepackage{float}
\usepackage{graphics}
\usepackage{grffile}
 \usepackage{pstool} 
\usepackage[textsize=tiny]{todonotes}
\usepackage{placeins}
\usepackage[ruled]{algorithm}                   % Algorithm environment
\usepackage{algorithmicx}                       % Algorithm environment
\usepackage[noend]{algpseudocode}                              % Pseudocode 

\usepackage{multirow}
\usepackage{stackengine}

\usepackage[square,comma,numbers]{natbib}
\usepackage[export]{adjustbox}

\usepackage{pgfplots}
\usepackage{pgfplotstable}
\usepackage{filecontents}

\usepackage{tikz}
\usepackage{tikzscale}
\usepackage{tikz-3dplot}

\usetikzlibrary{3d}
\usetikzlibrary{spy}
\usetikzlibrary{arrows,shapes}    
\usetikzlibrary{spy}
\usetikzlibrary{backgrounds}  
\usetikzlibrary{decorations}
\usetikzlibrary{decorations.markings}
\usetikzlibrary{positioning}
\usetikzlibrary{patterns}
\usetikzlibrary{calc} 
\usetikzlibrary{fadings}
\usetikzlibrary{trees}
\usepackage{booktabs}   
\usepackage{makecell} 
\usepackage{colortbl}   

\usepackage{xspace}
\usepackage{color}     
\usepackage{setspace}   
 
\usepackage{soul}
\usepackage{ulem}
\usepackage{currfile}
\usepackage{import}

\usepackage{authoraftertitle} % used for getting the authors 
\usepackage{authblk} % used for affiliations 

\pgfplotsset{compat=1.8}

\newcommand{\findmax}[3]{
    \pgfplotstablesort[sort key={#2},sort cmp={float >}]{\sorted}{#1}%
    \pgfplotstablegetelem{0}{#2}\of{\sorted}%
    \let #3=\pgfplotsretval%
}

\pgfplotscreateplotcyclelist{myCycleListBlackAndWhite}
{%
  {black, mark=o},
  {black, mark=square},
  {black, mark=triangle},
  {black, mark=diamond}, 
  {black, mark=pentagon}, 
  {black, mark=*},
  {black, mark=square*},
  {black, mark=triangle*},
  {black, mark=diamond*}, 
  {black, mark=pentagon*}, 
}

\definecolor{darkgreen}{rgb}{0,0.4,0} 
\definecolor{darkbrown}{rgb}{0.5, 0.396, 0.09}

\pgfplotscreateplotcyclelist{myCycleListColor}
{%
  {blue, mark=o},
  {red, mark=square},
  {darkbrown, mark=triangle},
  {darkgreen, mark=diamond},
  {black, mark=pentagon},
  {blue, mark=*},
  {red, mark=square*},
  {darkbrown, mark=triangle*},
  {darkgreen, mark=diamond*},
  {black, mark=pentagon*},
}       

\pgfplotsset{every axis/.append style= 
              {
                font=\small,
                mark size=2,
                line width = 0.1,
                legend style={font=\small, mark size=3, draw=none, fill=none},
                legend cell align=left,
                cycle list name=myCycleListColor,
              }
            } 
            
\pgfplotstableset
{
  every odd row/.style={before row={\rowcolor[gray]{0.9}}},
  every head row/.style=
  {
    before row=
    {%
      \toprule
    },
    after row=\midrule
  },
  every last row/.style=
  {
    after row=\bottomrule
  }
}

\newif\ifdrawboundingbox

\usetikzlibrary{external} % Is activated with the command
\tikzset{external/system call={pdflatex \tikzexternalcheckshellescape
-halt-on-error -interaction=batchmode -jobname "\image" "\texsource"}} 
  
\newcommand{\padnum}[2]{%
  \ifnum#1>1 \ifnum#2<10 0\fi
  \ifnum#1>2 \ifnum#2<100 0\fi
  \ifnum#1>3 \ifnum#2<1000 0\fi
  \ifnum#1>4 \ifnum#2<10000 0\fi
  \ifnum#1>5 \ifnum#2<100000 0\fi
  \ifnum#1>6 \ifnum#2<1000000 0\fi
  \ifnum#1>7 \ifnum#2<10000000 0\fi
  \ifnum#1>8 \ifnum#2<100000000 0\fi
  \ifnum#1>9 \ifnum#2<1000000000 0\fi
  \fi\fi\fi\fi\fi\fi\fi\fi\fi
  \expandafter\expandafter\number#2%
}

\newcounter{tmp}
\newcommand{\nextfigurename}[1]%
{%
	\setcounter{tmp}{\thefigure}%
	\addtocounter{tmp}{1}%
	\tikzsetnextfilename{section\thesubsection_#1}%
}% 

\newcommand{\nextsubfigurename}[1]%
{%
	\tikzsetnextfilename{section\thesubsection_#1}%
}%
          
\pgfkeys 
{ 
  /pgf/number format/.cd,
  fixed, 
  set thousands separator={\,},
  1000 sep in fractionals,
}  

\makeatletter
\renewcommand{\todo}[2][]{\tikzexternaldisable\@todo[#1]{#2}\tikzexternalenable}
\makeatother

\makeatletter
\renewcommand{\missingfigure}[2][]{\tikzexternaldisable\@missingfigure[#1]{#2}\tikzexternalenable}
\makeatother

\newcolumntype{C}[1]{>{\centering\arraybackslash}m{#1}}
\newcolumntype{R}[1]{>{\raggedright\arraybackslash}m{#1}}
\newcolumntype{L}[1]{>{\raggedleft\arraybackslash}m{#1}}

\newcommand{\includetikz}[2][]{%
\StrSubstitute{#2}{/}{_}[\temp]%
\tikzsetnextfilename{\temp}%
\includegraphics[#1]{#2.tikz}%
} 

\newcommand{\delete}[1]{\xspace} 
   
\newcommand{\rootDir}{}
\newcommand{\graphDir}{}
\newcommand{\dataDir}{}
\newcommand{\picsDir}{}

\newcommand{\halfTextWidth}{0.495\textwidth}

\title{Parallelization of the multi-level $hp$-adaptive finite cell
method}

\author[1]{John N. Jomo\thanks{john.jomo@tum.de, Corresponding
Author}}
\author[1]{Nils Zander}%\thanks{tino.bog@tum.de}}
\author[1]{Mohamed Elhaddad}%\thanks{tino.bog@tum.de}}
\author[1]{Ali \"Ozcan}%\thanks{kollmannsberger@mytum.de}}
\author[1]{\authorcr Stefan Kollmannsberger}%\thanks{kollmannsberger@mytum.de}}
\author[1]{Ralf-Peter Mundani}%\thanks{kollmannsberger@mytum.de}}
\author[1/2]{Ernst Rank}%\thanks{ernst.rank@tum.de}}

\affil[1]{Chair for Computation in Engineering,
 		  Technische Universit\"at M\"unchen,
  		  Arcisstr. 21, 80333 M\"unchen, Germany}
\affil[2]{Institute for Advanced Study, 
	  Technische Universit\"at M\"unchen, Lichtenbergstr. 2a, 85748 Garching,
	  Germany}

\usepackage{fancyhdr}
\usepackage{hyperref}
\date{}
\fancypagestyle{plain}{%
  \fancyhf{}%
  \fancyfoot[L]{
  \vspace{-1.25cm} 
  \footnotesize{\copyright 2017. Licensed under the Creative Commons CC-BY-NC-ND 4.0 license\\ \url{http://creativecommons.org/licenses/by-nc-nd/4.0/}
  }} }

\drawboundingboxtrue    
\drawboundingboxfalse     

\begin{document}      
\normalem          
\maketitle             
    
%% Abstract ---------------------------------------
\vspace{-1.5cm} 
\hrule 
\section*{Abstract}
%\todo{write abstract}
The multi-level $hp$-refinement scheme is a powerful extension of the finite element method that allows local mesh adaptation without the trouble of constraining hanging nodes. This is achieved through hierarchical high-order overlay meshes, a $hp$-scheme based on spatial refinement by superposition. An efficient parallelization of this method using standard domain decomposition approaches in combination with ghost elements faces the challenge of a large basis function support resulting from the overlay structure and is in many cases not feasible. In this contribution, a parallelization strategy for the multi-level $hp$-scheme is presented that is adapted to the scheme's simple hierarchical structure. By distributing the computational domain among processes on the granularity of the active leaf elements and utilizing shared mesh data structures, good parallel performance is achieved, as redundant computations on ghost elements are avoided. We show the scheme's parallel scalability for problems with a few hundred elements per process. Furthermore, the scheme is used in conjunction with the finite cell method to perform numerical simulations on domains of complex shape.

%% Keywords ---------------------------------------
\vspace{0.25cm}
\noindent \textit{Keywords:} high-order FEM, automatic $hp$-adaptivity, arbitrary hanging
nodes, finite cell method, parallel computation, high performance computing
 
\vspace{0.25cm}
\hrule  

% \tableofcontents 
 
%% Actual Content ---------------------------------
\section{Introduction}
\paragraph{}
The recently proposed multi-level $hp$-refinement \cite{Zander2015} is a novel approach for performing adaptive mesh refinement in the context of high-order finite elements. Contrary to conventional $hp$-adaptive methods, which perform spatial refinement by the replacement of existing elements with a subset of smaller elements \cite{Solin2004a,BangerthHartmannKanschat2007,Demkowicz2007}, this method employs hierarchical, high-order overlay meshes to improve the quality of the approximation in areas of interest. The main advantages of this approach over conventional ones are its ability to deal with an arbitrary number of hanging nodes and its intuitive refinement and coarsening procedures that make it easy to implement and suitable for performing dynamic mesh refinements. Moreover, this $hp$-scheme maintains the same exponential convergence rates characteristic of classical $hp$-formulations even in the presence of singularities and is easily extensible to three dimensions \cite{Zander2016}. 

As shown in \cite{Zander2015}, multi-level $hp$-adaptivity can be combined with the finite cell method to solve partial differential equations on complex domains. The finite cell method, introduced by Parvizian \textit{et al.} \cite{Parvizian2007}, is a high-order fictitious domain approach that circumvents the tedious process of mesh generation by embedding the physical domain in a computational domain that can be trivially meshed. Computations on complex domains are hereby simplified as the geometrical information needs only to be resolved on integration level. The combination of the finite cell method and multi-level $hp$-adaptivity, coined the multi-level $hp$-adaptive finite cell method, is currently being applied in the calculation of engineering problems in the fields of biomechanics and additive manufacturing.

Parallel computations on modern computing systems using finite element techniques have made the simulation of large engineering problems possible. A plethora of distributed memory parallelization strategies have been developed for different finite element discretization schemes such as \cite{Paszynski2006,paszynski_2011,bangerth_2008}. These strategies not only aim at reducing computational time but also at efficiently utilizing the available hardware resources. The major challenge in developing such strategies is the choice of algorithms and data structures that best expose the parallelism in the discretization scheme under consideration. As a consequence, different studies have been undertaken such as in \cite{BangerthHartmannKanschat2007,Popescu_2013}, which seek to provide guidelines for the parallelization of finite element schemes.    

One central algorithm in every parallelization scheme is the distribution of the computational domain among participating processes. A common approach motivated by non-overlapping domain decomposition techniques entails subdividing the computational domain into sub-domains of fairly equal computational cost and their subsequent distribution among the processes. This distribution, in the context of parallel $hp$-adaptive schemes, can be first performed on the granularity of the initial elements \cite{Paszynski2006}, followed by a re-balancing step of the refined grid or directly on the refined grid. Each process stores its local elements as well as so called ghost elements that constitute the boundary to neighboring sub-domains \cite{sampath2010parallel}. This approach is advantageous in maintaining mesh compatibility after parallel mesh refinement \cite{Paszynski2006} and more importantly when setting up the global linear system to be solved. This system can be assembled without communication since each process can compute the global contributions of the degrees of freedom it owns by integrating the corresponding local elements and ghost elements \cite{sampath2010parallel,Popescu_2013}. Different variants of the described domain decomposition approach have been successfully applied to $hp$-adaptive schemes, as in \cite{BangerthHartmannKanschat2007,Paszynski2006,mfem-library}, and can be combined with distributed mesh data structures to reduce the amount of memory needed by a single process.

Although ghost elements eliminate the need for communication during distributed assembly, redundant computations are performed on ghost elements. Consequently, the ratio of local to ghost elements on a process has to be high, in order to hide redundant computations and allow scalability \cite{BangerthHartmannKanschat2007}. Usage of the ghost element approach in the parallelization of the multi-level $hp$-scheme would, however, result in a significantly larger number of ghost elements than in conventional $hp$-schemes, thus inhibiting scalability. This is due to the large support of the basis functions, as basis functions are coupled over the hierarchy of the overlay meshes. Moreover, the number of basis functions and consequently the computational work vary highly from element to element in multi-level $hp$-refinement, further complicating load balancing. 

This contribution presents a parallelization strategy for the multi-level $hp$-refinement that addresses the mentioned challenges. Our scheme avoids the use of ghost elements by utilizing a shared mesh data structure, a concept first applied to a conventional $hp$-scheme in \cite{paszynski_2011}. This method is adopted to take advantage of the simplicity of multi-level $hp$-refinement.  The absence of ghost elements results in the need for data communication during the assembly of the global linear system. This, however, does not significantly affect parallel performance as shown in the numerical examples considered. Moreover, since no redundant computations are performed during integration, good parallel performance for problems with a few hundred elements per process is achieved. The present work also shows the combination of the parallel multi-level $hp$-scheme with the finite cell method for the computation of engineering problems involving complex geometries. 

This paper is organized as follows: The main ideas behind multi-level $hp$-refinement and the finite cell method will be outlined in Section \ref{multiLevelFCM}. This will be followed by a description of the central aspects of the parallel implementation in Section \ref{parallelImplementation}. Numerical examples in Section \ref{numericalResults} show the parallel performance of the proposed method for different problem classes. Finally, Section \ref{conclusion} will conclude the contribution with a brief summary and an outlook into future aspects of research.

\section{The multi-level $hp$-adaptive finite cell method}\label{multiLevelFCM}

\renewcommand{\rootDir}{multiLevelFCM}
\renewcommand{\picsDir}{\rootDir/pics}

A significant improvement of the approximation quality in finite element simulations can be achieved by the use of $hp$-adaptive discretizations which perform local mesh refinement in areas of interest. This section provides a description of such a method, the multi-level $hp$-refinement, focusing on the characteristics that enable it to leverage the benefits of $hp$-adaptive methods while notably reducing the implementational effort. The basic idea behind the finite cell method is presented and the combined use of both methods in the simulation of complex problems is also highlighted.   

\subsection{Multi-level $hp$-refinement}
Although $hp$-adaptive formulations posses superior approximation qualities, their implementational complexity inhibits the widespread use of these methods in the field of computational mechanics. This complexity stems mainly from the algorithms and data structures needed to deal with mesh irregularities that arise during refinement and coarsening procedures. Classical $hp$-formulations commonly perform mesh refinement by replacing elements having a high discretization error with a set of smaller elements \textit{cf.} \cite{Solin2004a,Paszynski2006}. This process introduces mesh irregularities, usually referred to as hanging nodes, as the basis functions of the new elements lack a corresponding counterpart in their unrefined neighboring elements \cite{hughes2000finite}. Special algorithms are therefore needed to constrain these hanging nodes and restore inter-element continuity. Many $hp$-formulations, as a consequence, only allow one level of mesh irregularity between elements e.g.\,\cite{Demkowicz2007,Paszynski2006}, while others implement sophisticated constraining algorithms that can deal with arbitrary levels of hanging nodes such as \cite{Solin2008,Solin2010a}.

Zander \textit{et al.} propose a novel $hp$-adaptive approach in \cite{Zander2015} based on refinement by superposition that circumvents the difficulties introduced by hanging nodes. By utilizing simple rule-sets to ensure compatibility and linear independence of the shape functions, intuitive refinement and coarsening procedures are developed. The need for complex data structures and algorithms to consolidate and constrain hanging nodes is alleviated as hanging nodes are avoided by construction. The approach is shown to achieve exponential convergence even in the presence of singularities and was first introduced for the two-dimensional case, extended in \cite{Zander2016} to three dimensions and applied to cohesive fracture modeling in \cite{Zander2016a}.

\subsubsection{Basic idea}\label{basicIdeaMLhp}
The core idea of the multi-level $hp$-approach is to locally superpose coarse base elements with finer overlay elements that better capture the solution characteristics in an area of interest. The final solution is hence the sum of the large scale solution $\boldsymbol{u}_{b}$ on the high-order coarse base mesh and the fine-scale solution $\boldsymbol{u}_{o}$ from the finer overlay mesh: $\boldsymbol{u} = \boldsymbol{u}_{b} + \boldsymbol{u}_{o}$. This superposition approach is attributed to the work of Mote in 1971 \cite{mote_global-local_1971} and has been adapted in different approaches e.g.  \cite{belytschko_spectral_1990,Rank1992,moore_adaptive_1992}.

\begin{figure}[t]
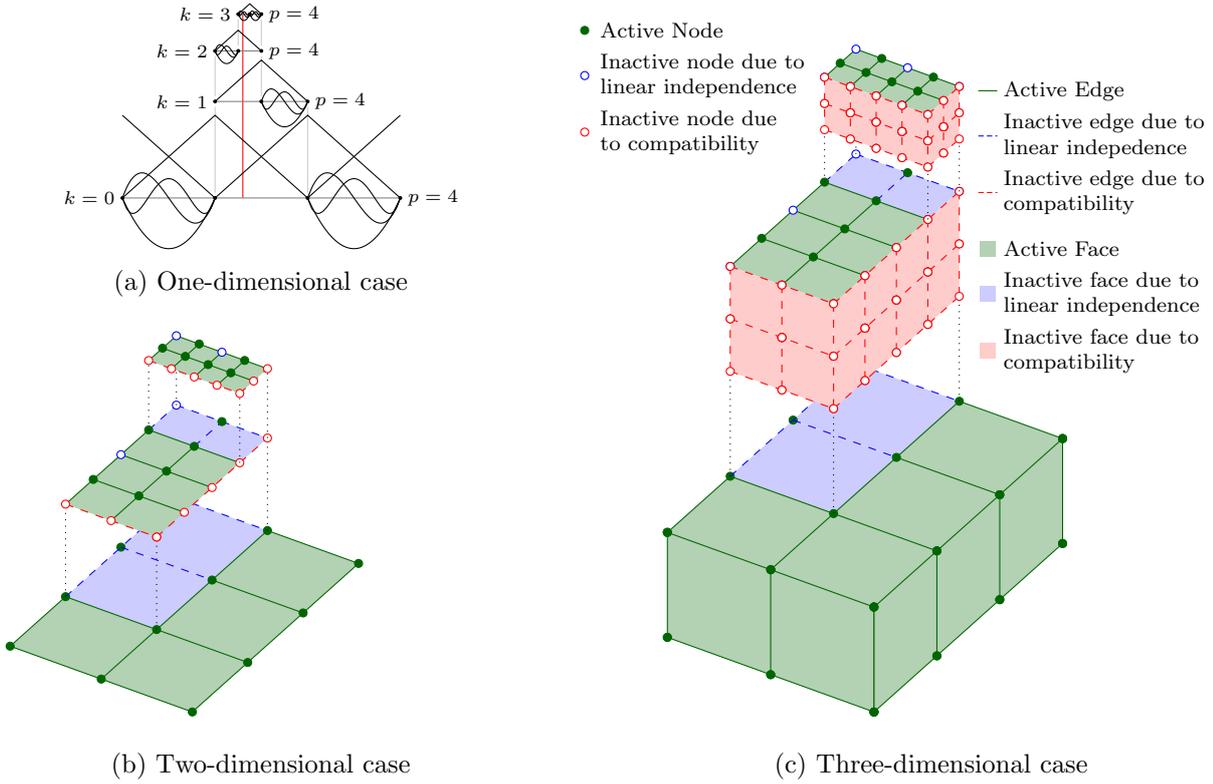
%
  \begin{minipage}[b]{.4\textwidth}
    \begin{center}
    \includetikz[width=0.8\textwidth]{\picsDir/refinement_1d_mlhp_woNumbers}%
    \end{center}
    \begin{center}
      \small
      \vspace{-0.3cm}
      (a) One-dimensional case
    \end{center}
    \includetikz[width=0.7\textwidth]{\picsDir/multiLevelhp2d}%
    \begin{center}
      \small
      %\vspace{-0.8cm}
      (b) Two-dimensional case
    \end{center}
  \end{minipage}%
  \hfill%
  \begin{minipage}[b]{.55\textwidth}
    \nextfigurename{overlay3d}%
    \includetikz[width=0.9\textwidth]{\picsDir/multiLevelhp3d}%
    \begin{center}
        \small
        (c) Three-dimensional case
      \end{center}
  \end{minipage}
  \caption{Basic idea of the multi-level $hp$-method \cite{Zander2016}.}
  \label{fig::multilevelhp::multilevelhpIdea}%
\end{figure}

Multi-level $hp$-refinement extends the original superposition approach through recursive superposition resulting in multiple levels of hierarchical overlay meshes. Furthermore, integrated Legendre shape functions are used to span the approximation space. The direct association of these shape functions with the element topology (nodes, edges, faces and solids) allows the creation of simple mesh refinement and coarsening techniques as individual basis functions can be easily eliminated from the ansatz space through deactivation of the corresponding topological entities. This important property enables the maintenance of mesh compatibility between adjacent elements and linear independence across the overlay meshes. Inter-element continuity is ensured by construction through the application of homogeneous Dirichlet boundary conditions on the overlay mesh. Consequently, arbitrary levels of hanging nodes can be used. $C^{\infty}$-continuity is thus ensured within each element and $C^{0}$-continuity across element boundaries. Linear independence, on the other hand, is guaranteed by the deactivation of topological components in the overlay meshes with active sub-components. 
Figure \ref{fig::multilevelhp::multilevelhpIdea} illustrates multi-level $hp$-refinement for the one, two and three-dimensional case. A multi-level $hp$-mesh consists of three different kinds of elements, firstly elements on the lowest level, so called base elements, secondly refined elements on intermediate refinement levels and finally elements on the highest refinement level with no sub-elements, so called leaf elements. Elements with no sub-elements are referred to as active leaf elements or in short active elements, such elements are either base or leaf elements.  Linear independence and compatibility of all basis functions over this element-hierarchy is achieved through the deactivation of topological components as shown in Figure \ref{fig::multilevelhp::multilevelhpIdea}. The interested reader is referred to \cite{Zander2015,Zander2016} for a comprehensive description of the refinement and coarsening strategies.
\subsubsection{Numerical integration in multi-level $hp$-refinement}\label{numericalIntegration}
\begin{figure}[t]
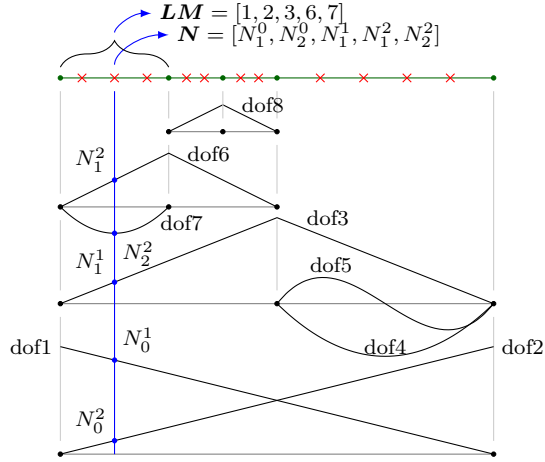
%
  \begin{center}%
  \includetikz[width=0.43\textwidth]{\picsDir/refinement_1d_mlhp_graded}%
  \end{center}%
  \caption{Integration of a multi-level element by composed Gaussian integration \cite{Zander2016}.}%
  \label{fig::multiLevelFCM::integration}%
\end{figure}

An important aspect of multi-level $hp$-refinement is the correct numerical integration of the element matrices. Applying conventional Gaussian quadrature at base element level would yield wrong results since the shape functions are only $C^{0}$-continuous within the base elements due to the hierarchical superposition. This problem is solved in \cite{Zander2016} by evaluating integrals separately on integration domains in which the basis functions are $C^{\infty}$-continuous. These integration domains are obtained by a projection of the leaf elements through the mesh hierarchy onto the base elements as illustrated in Figure \ref{fig::multiLevelFCM::integration} for the one dimensional case. Numerical integration can then be performed independently on each integration domain as shown. Figure \ref{fig::multiLevelFCM::integration} also shows the coupling of the basis functions over the different mesh levels. This property of the multi-level $hp$-scheme does not affect performance, as non-zero shape functions within an integration domain can be efficiently computed by concatenating the active degrees of freedom over the different mesh levels. 
\subsection{The finite cell method}
\begin{figure}[H]
  \begin{center}
      %\tikzset{external/remake next}%
      \nextfigurename{fcm}%
      \includegraphics[width=0.95\textwidth]
        {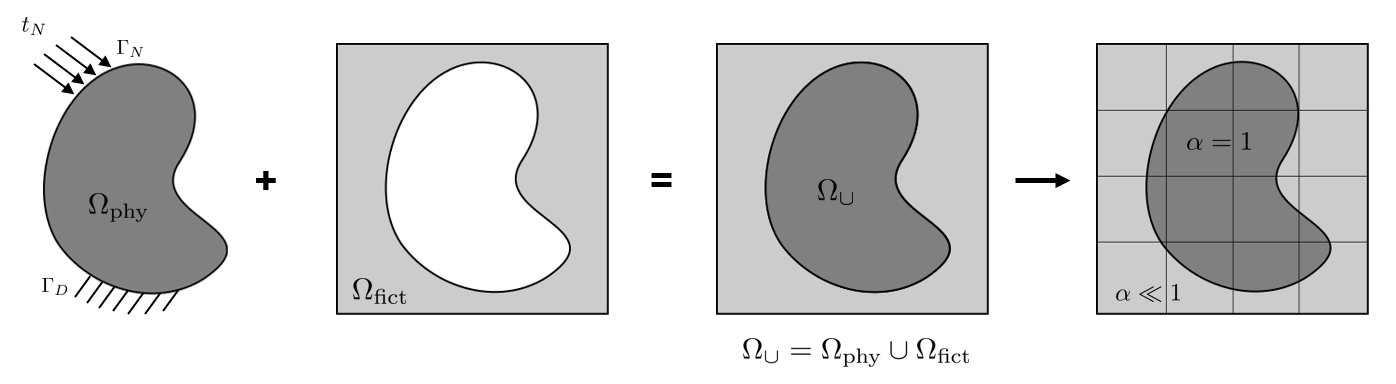}%
  \end{center}
  \caption{Basic idea behind the finite cell method following \cite{schillinger_small_2012}.}
  \label{fig::multiLevelFCM::fcm}
\end{figure}

The aforementioned finite cell method aims to circumvent the tedious process of mesh generation for complex geometries by combining a fictitious domain approach with high-order finite elements. The physical domain $\Omega_{\text{phy}}$ is extended by a fictitious domain $\Omega_{\text{fict}}$, yielding a computational domain $\Omega_{\cup}$ from the union $\Omega_{\text{phy}} \cup \Omega_{\text{fict}}$, which can be easily meshed using high-order finite elements on a structured grid as illustrated in Figure \ref{fig::multiLevelFCM::fcm}. The elements are referred to as finite cells to differentiate them from standard boundary conforming finite elements \cite{Duster2008}. The original geometry is resolved at integration level by means of an indicator function $\alpha(\boldsymbol{x})$ 
\begin{equation}\label{eq::multiLevelFCM::alpha}
	\alpha(\boldsymbol{x}) = \bigg\{ \begin{matrix}
			\ 1&&  \forall \, \boldsymbol{x} \in \ \Omega_{\text{phy}}
			\\ \varepsilon \ll 1&& \forall \, \boldsymbol{x} \in \ \Omega_{\text{fict}}
	\end{matrix} \ , 
\end{equation}
which associates a given point $\boldsymbol{x}$ with the physical or fictitious domain. This results in a modified weak form of the governing differential equations, which is now scaled by the scalar field $\alpha(\boldsymbol{x})$. To illustrate this, we consider the weak form of the equilibrium equation in linear elasticity given as $\mathcal{B}(\boldsymbol{u},\boldsymbol{v}) = \mathcal{F}(\boldsymbol{v})$. The terms $\boldsymbol{u}$ and $\boldsymbol{v}$ represent the displacement field and test functions respectively, while $\mathcal{B}(\boldsymbol{u},\boldsymbol{v})$ represents the bilinear form 
which is calculated from the linear strain operator $\mathbf{B}$ and material matrix $\mathbf{C}$. The term $\mathcal{F}(\boldsymbol{v})$ is the linear functional representing the contributions of the volume loads $\boldsymbol{f}$ and tractions $\boldsymbol{t}_N$ acting on the Neumann boundary $\Gamma_{N}$. 
\begin{equation}
	\mathcal{B}(\boldsymbol{u},\boldsymbol{v}) = \int \limits_{\Omega } \mathbf{B}^{T} \, \mathbf{C} \, \mathbf{B} \, \text{d}\Omega \ \ \  \text{and} \ \ \
 \mathcal{F}(\boldsymbol{v}) = \int \limits_{\Omega} \boldsymbol{v}^{T} \boldsymbol{f} \, \text{d}\Omega \ +  \int \limits_{\Gamma_{N}} \boldsymbol{v}^{T} \boldsymbol{t}_{N} \, \text{d}\Gamma \ .
\end{equation}
%
%\begin{equation}\label{eq::multiLevelFCM::linearF}
%\mathcal{F}(\boldsymbol{v}) = \int \limits_{\Omega} \boldsymbol{v}^{T} \boldsymbol{f} \, \text{d}\Omega \ +  \int \limits_{\Gamma_{N}} \boldsymbol{v}^{T} \boldsymbol{t}_{N} \, \text{d}\Gamma \ .
%\end{equation}
%
The indicator function $\alpha(\boldsymbol{x})$ introduced in FCM leads to a modified bilinear form given as
\begin{eqnarray}\label{eq::multiLevelFCM::integration}
	\bar{\mathcal{B}}(\boldsymbol{u},\boldsymbol{v}) = \int \limits_{\Omega_{\cup} } \mathbf{B}^{T} \alpha(\boldsymbol{x}) \, \mathbf{C} \, \mathbf{B} \, \text{d}\Omega   
      \  = \int \limits_{\Omega_{\text{phy} } } \mathbf{B}^{T} \, 1 \, \mathbf{C} \, \mathbf{B} \, \text{d}\Omega \ + \int \limits_{\Omega_{\text{fict}}} \mathbf{B}^{T} \, \varepsilon \, \mathbf{C} \, \mathbf{B} \, \text{d}\Omega \ . 
\end{eqnarray}
As shown in \cite{Dauge2015}, the numerical approximation resulting from the modified bilinear form and the linear functional in Equation \ref{eq::multiLevelFCM::integration} converges to the solution of the original weak form when epsilon tends to zero.
%\bigbreak \noindent
The extended domain in FCM typically has a very simple geometric structure rendering it easy to mesh. In the simplest case, a uniform grid of rectangular hexahedral grids is used, yet the hierarchical overlay meshes described in the Section \ref{basicIdeaMLhp} are also applicable.

The scalar field $\alpha(\boldsymbol{x})$ introduces a discontinuity within elements that are intersected by the boundary of the physical domain. Different approaches have been developed to improve the integration of these cut elements by approximating the original domain boundary such as a low order approximation technique utilizing composed Gaussian quadrature in combination with recursive spacetree-subdivision \cite{Duster2008}, as well as high-order techniques like moment fitting \cite{Joulaian2016} and the recently proposed blended partitioning algorithm \cite{Kudela2015,Kudela2016}. 

Another important aspect of the finite cell method is the imposition of boundary conditions. While the application of Neumann boundary conditions remains unaffected by a nonconforming discretization, essential boundary conditions are. The latter are usually imposed weakly using variational techniques like the penalty method \cite{Babuska1973,Fernandez-Mendez2004}, Nitsche's Method \cite{Nitsche1971,Fernandez-Mendez2004} and the Lagrange multiplier method \cite{ruess_weakly_2013}.

The finite cell method maintains the excellent approximation qualities of standard $p$ finite elements, while significantly simplifying the mesh generation process. It has thus been successfully applied to different application fields which including linear elasticity \cite{Parvizian2007,Duster2008}, large deformation analysis \cite{schillinger_small_2012}, image-based simulation of bones \cite{Verhoosel2015}, contact mechanics \cite{Bog2015} and isogeometric analysis \cite{schillinger_isogeometric_2012} among others. The interested reader can refer to \cite{Parvizian2007,Duster2008,Schillinger2014} for an in depth presentation of FCM or an open-source \texttt{Matlab} implementation \cite{Zander2014} for a simple introduction into this research field. A combination of multi-level $hp$-refinement and FCM yields a highly flexible discretization method which can be used to perform static and transient simulations on complex geometries. In multi-level $hp$-FCM, the composed integration techniques are applied on the integration domains described in Section \ref{numericalIntegration}. Ongoing research aims to exploit the benefits of both methods in biomechanical simulation of bone-structures and in the simulation of the additive manufacturing process.

\section{Parallel Implementation}\label{parallelImplementation}
\renewcommand{\rootDir}{parallelImplementation}
\renewcommand{\graphDir}{\rootDir/graphs/}
\renewcommand{\dataDir}{\rootDir/data/}
\renewcommand{\picsDir}{\rootDir/pics}
Large numerical simulations running on distributed memory systems greatly benefit from the ability of $hp$-formulations to simultaneously resolve small and large geometrical scales with a reasonable number of unknowns. Such methods have been successfully applied to multi-scale problems e.g. in simulating seismic waves \cite{Breuer_2014} or mantel convection \cite{Ruede_2015}, simulations in which application of uniform global $h$-refinement would not be feasible. The parallelization of $hp$-formulations has caught the interest of many research groups and lead to the emergence of different strategies \cite{paszynski_2011,Paszynski2006} and various open source parallel $hp$-adaptive frameworks like \texttt{deal.II} \cite{BangerthHartmannKanschat2007}, \texttt{Hermes} \cite{Hermes-project}, \texttt{libMesh} \cite{libMeshPaper}, \texttt{Nektar++} \cite{Cantwell2015205} and \texttt{MFEM} \cite{mfem-library} with a wide user base in the scientific community. The majority of the mentioned parallel $hp$-codes use classical $hp$-formulations that perform refinement by replacement. Moreover, several software projects like \texttt{Trilinos} \cite{Trilinos} and \texttt{PETSc} \cite{petsc-efficient} have emerged, which provide different functionalities in the field of computer science, e.g parallel linear algebra packages.   
\subsection{Parallelization strategy}
\begin{figure}[!htb]
\centering
\begin{minipage}{0.46\textwidth}
  \centering	
  \begin{minipage}{0.28\textheight}
   \centering	  
      \includegraphics[width=0.5\textwidth]{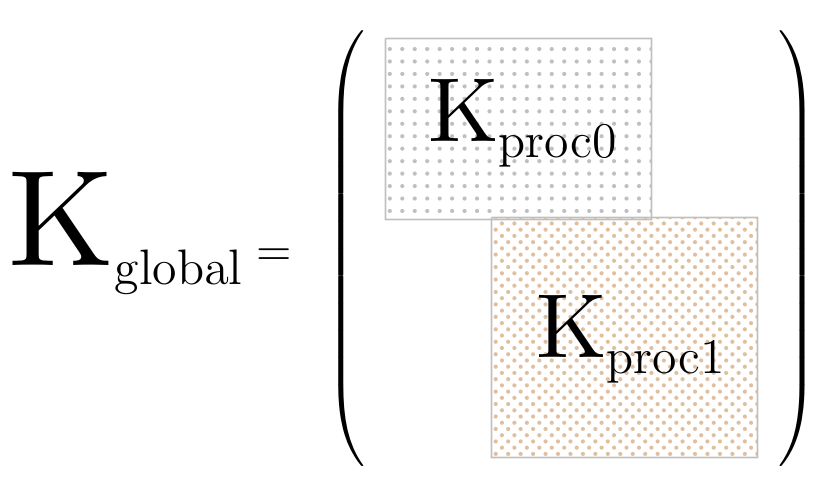}%
  \end{minipage}
  \begin{minipage}{0.7\textheight}
	  \stackunder[5pt]{\includetikz[width=0.25\textwidth]{\picsDir/normal-hp2-mod}}{\small view process 0}%
	  \stackunder[5pt]{\includetikz[width=0.25\textwidth]{\picsDir/normal-hp1-mod}}{\small view process 1}%
  \end{minipage}
  \small {\color{white} create white line to keep alignment} \\
  \small (a) Domain decomposition with ghost elements.
\end{minipage}
\hspace{8mm}
\begin{minipage}{0.46\textwidth}
\centering
  \begin{minipage}{0.28\textheight}
   \centering	  
      \includegraphics[width=0.5\textwidth]{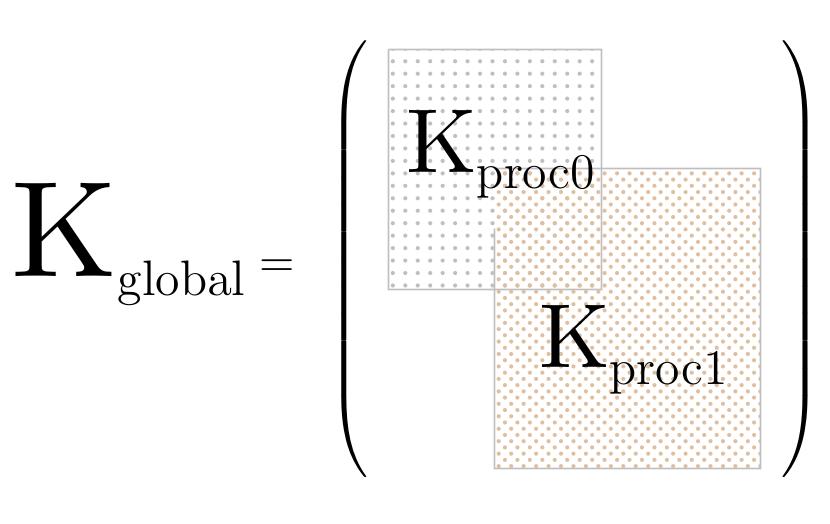}%
  \end{minipage}
  \begin{minipage}{0.7\textheight}
	  \stackunder[5pt]{\includetikz[width=0.25\textwidth]{\picsDir/activemlhp2}}{\small view process 0}%
	  \stackunder[5pt]{\includetikz[width=0.25\textwidth]{\picsDir/activemlhp1}}{\small view process 1}%
  \end{minipage}
  \small {\color{white} create white line to keep alignment} \\
  \small (b) Domain decomposition without ghost elements.
\end{minipage}
\hfill
\caption{Comparison of two domain decomposition strategies.}
\label{fig::parallelImplementation::comparisonDomainDecomposition}%
\end{figure}

The domain decomposition approach with ghost elements mentioned in the introduction is commonly used when distributing the computational domain among processes \cite{Popescu_2013,sampath2010parallel,mfem-library}. Although this approach has different variants, the main idea is that each process stores elements on its portion of the original domain and ghost elements that constitute the boundary to neighboring sub-domains. In this way, communication during the assembly of the linear system is avoided as illustrated in Figure \ref{fig::parallelImplementation::comparisonDomainDecomposition}a. The degrees of freedom on each process consist of entries local to the process and shared entries due to the ghost elements. Each process only computes the full contribution of its local degrees of freedom by integrating the associated local elements and ghost elements represented by the green and dashed elements in Figure \ref{fig::parallelImplementation::comparisonDomainDecomposition}a, respectively.

\begin{figure}[t]
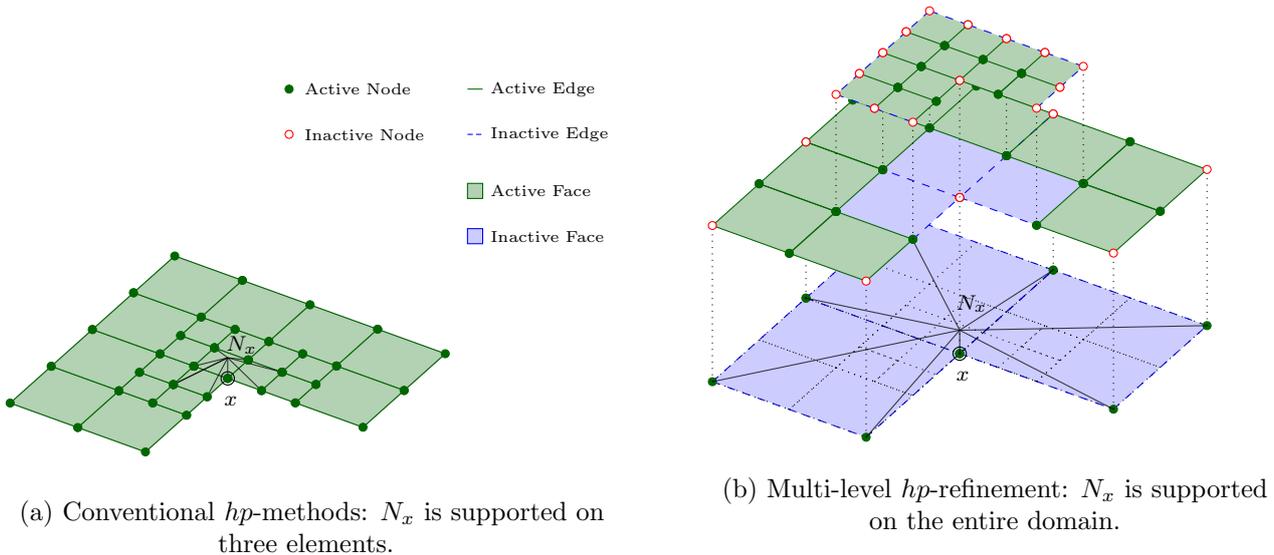
%
  \begin{minipage}[b]{.48\textwidth}
    \begin{center}
    \includetikz[width=1.0\textwidth]{\picsDir/normal-hp}%
    \end{center}
    \begin{center}
      \small
        (a) Conventional $hp$-methods: $N_{x}$ is supported on three elements.
      \vspace{-0.3cm}
    \end{center}
  \end{minipage}%
  \hfill%
  \begin{minipage}[b]{.45\textwidth}
    \nextfigurename{overlay3d}%
        \includetikz[width=1.0\textwidth]{\picsDir/new-modified-multiLevelhpDomainDecomp}%
    \begin{center}
        \small
      (b) Multi-level $hp$-refinement: $N_{x}$ is supported on the entire domain.
      \end{center}
  \end{minipage}
  \caption{Comparison of the basis function support in multi-level $hp$-refinement and conventional $hp$-methods.}
  \label{fig::parallelImplementation::discussionDomainDecomp}%
\end{figure}

A parallel implementation utilizing the domain decomposition approach with ghost elements is not well suited for multi-level $hp$-refinement. This is due to the large support of the basis functions as stated in the introduction. To illustrate this point, an L-shaped domain comprising of three base elements is refined in two steps towards the reentrant corner. The mesh resulting from multi-level $hp$-refinement is compared to a mesh obtained from a conventional $hp$-scheme in Figure \ref{fig::parallelImplementation::discussionDomainDecomp}. In order to compute the entries in the linear system associated with basis function $N_x$ at node $x$, a total of three elements need to be integrated in the conventional $hp$-scheme. This is, however, not the case in the multi-level $hp$-scheme. The basis function $N_x$ is supported on all 21 integration domains, represented by the dashed lines in the base elements in Figure \ref{fig::parallelImplementation::discussionDomainDecomp}b.  These domains have to be integrated in order to compute the entries in the linear system associated with $N_x$ without communication as in Figure \ref{fig::parallelImplementation::comparisonDomainDecomposition}a. We can safely conclude that no speed up can be attained in this particular example. This example illustrates the shortcomings of the conventional ghost element parallelization strategy when applied to multi-level $hp$-refinement, and further motivates the development of a parallelization scheme that is better adapted to multi-level $hp$-refinement. 

We propose a scheme based on the distribution of the integration domains described in Figure \ref{fig::multiLevelFCM::integration} in order to guarantee scalability even for rather coarse meshes, such as in Figure \ref{fig::parallelImplementation::discussionDomainDecomp}b. A shared mesh data structure, similar to \cite{paszynski_2011}, is employed due to its low implementational effort while at the same time allowing for an easy activation and deactivation of elements during refinement. Mesh concurrency can thus be easily maintained among processes during a simulation. The global linear system is assembled with communication as illustrated in Figure \ref{fig::parallelImplementation::comparisonDomainDecomposition}b. The individual components of the scheme shall now be described in the following sections.

\subsection{Algorithms and data structures}
Our parallelization strategy is implemented within an object-oriented framework and performs partitioning of the computational domain on the granularity of the integration domains, an approach specially crafted for multi-level $hp$-refinement. This can also be interpreted as a distribution of the active elements, due to the direct association between integration domains and active elements, see Figures \ref{fig::multilevelhp::multilevelhpIdea} and \ref{fig::multiLevelFCM::integration}. The strategy consists of a serial part that involves mesh initialization and refinement followed by a parallel part that constitutes mesh partitioning, integration of the linear system and its distributed assembly, solving of the linear system and post processing. A summary of the simulation pipeline is given in Algorithm \ref{alg:simulation_pipeline}.

\subsubsection{Mesh initialization and refinement}
The initial computational domain is generated by every process at simulation start. Adaptive mesh refinement using the multi-level $hp$-scheme is performed redundantly by each process on the whole domain, resulting in the same discretization on all processes. This approach is adopted as multi-level $hp$-refinement is currently driven by a priori information. Mesh refinement, therefore, does not need to be divided among the processes, but can efficiently performed by each process, since the elements to be refined are known beforehand. Furthermore, a priori refinement allows mesh partitioning to be directly performed on the refined grid.  Multi-level $hp$-refinement can, however, be extended to perform refinement driven by error indicators and estimators, as shown in \cite{DiStolfo2016}, and is a topic of ongoing research \cite{DAngella2016}. Such automatic $hp$-refinement would, however, require more involved fully parallel refinement strategies and data structures such as those described in \cite{Paszynski2006,BangerthHartmannKanschat2007}, which perform a distribution of the initial elements, followed by parallel refinement on a subset of elements and a subsequent re-balancing of the refined grid on the granularity of either initial elements or refined elements in order to guarantee scalability. 

\begin{figure*}[h]
  \begin{center}
     %\tikzset{external/remake next}%
     \includegraphics[width=0.85\textwidth]{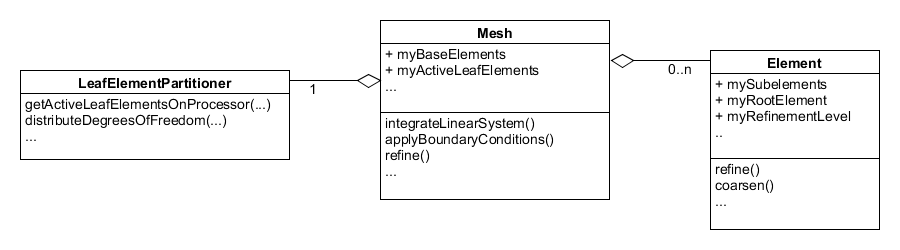}%
  \end{center}
  \caption{ Mesh data structure in the parallel multi-level $hp$-refinement.}
  \label{fig::parallelImplementation::meshdataStructure}
\end{figure*}

\begin{figure*}[h]
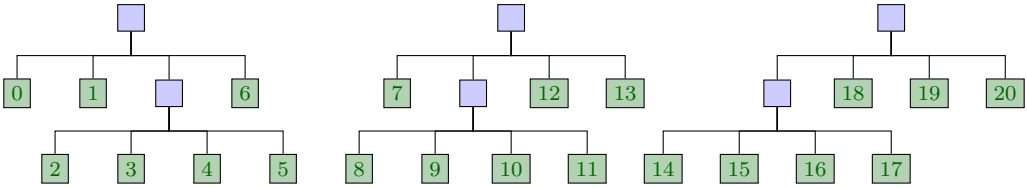

  \begin{center}
     %\tikzset{external/remake next}%
     \includetikz[width=0.8\textwidth]{\picsDir/tree}%
  \end{center}
  \caption{Implicit tree-structure of the L-shaped domain mesh in multi-level $hp$-refinement showing the numbering of the active leaf elements.}
  \label{fig::parallelImplementation::meshImplicitGraph}
\end{figure*}

The aforementioned simplicity of the shared mesh data structure coupled with a priori $hp$-refinement, yields a fast, easy to implement $hp$-scheme as shown by the numerical examples in Section \ref{numericalResults}. The data structures used are implemented in terms of pointers as described in \cite{Zander2015,Zander2016a}, where the mesh stores elements as pointers. Mesh elements in turn, hold pointers to their sub-elements/children, allowing for easy navigation through the mesh hierarchy. This structure is portrayed in a simple UML-diagram in Figure \ref{fig::parallelImplementation::meshdataStructure}. It is important to note that the refinement yields an implicit tree-structure, which is defined through the different pointer relations as shown in Figure \ref{fig::parallelImplementation::meshImplicitGraph}.

\subsubsection{Load balancing: Distribution of the integration domains} \label{distribution_leafelements}
Upon mesh refinement, a computational domain consisting of elements with different refinement levels is obtained. The leaf elements can now be partitioned among processes, a task performed by the \textit{LeafElementPartitioner} illustrated in Figure \ref{fig::parallelImplementation::meshdataStructure}. This class is aggregated by the mesh and performs load balancing after every mesh refinement step by first creating a trivial intial partitioning, where each process is assigned a set of contiguous leaf elements. The \textit{LeafElementPartitioner} also provides an interface to the geometric and graph-based partitioners in \texttt{Zoltan} \cite{ZoltanShortTutorial}.  The initial partitioning created in the first partitioning step, alongside auxiliary information such as leaf element weights, are forwarded to the \textit{LB\_Partition} function in \texttt{Zoltan}. This function improves the initial leaf element distribution in a second step, yielding a set of unique active leaf elements $\Omega^{h}_{proc}$ on every process, which constitutes the final leaf element distribution. The described load balancing scheme is summarized in Algorithm \ref{alg:distribution_of_integration_domains}.  This scheme greatly benefits from the use of a shared data mesh structure, as all operations are based solely on the indices's of the leaf elements and auxiliary geometrical information computed by each processor on a unique subset of elements (initial partitioning). Furthermore, no elements need to be packed and sent to other processors. Communication occurs only within the \textit{LB\_Partition} function in \texttt{Zoltan}, which returns a set of unique indices on each processor. The \textit{LeafElementPartitioner} registers these indices to the mesh on each individual process, marking the respective leaf elements as active leaf elements that can be processed in the subsequent integration.

\begin{algorithm}[h!]
	\caption{ Load balancing : Distribution of integration domains (leaf elements) }
\label{alg:distribution_of_integration_domains}
\begin{algorithmic}[1]
	\Function{\texttt{get\_active\_leafelements\_onprocess} }{ mesh }
    \State leafIndexVector = \texttt{assignUniqueIndicesToLeafElements}( mesh.baseElements )
    \State myProcessLeafIds = \texttt{assignContiguousLeafElementsToProcess}( leafIndexVector )
%    \Statex
    \State \Comment{Now improve the initial distribution of leaf elements using Zoltan }
    \If { ( partitioningType == geometric ) }
    \State myLeafCentroids, weights = \texttt{getCentroidsAndWeights}( myProcessLeafIds, mesh )
	\State myProcessLeafIds = \texttt{improveLoadBalance}( myLeafCentroids, weights )
	\State \Comment{call to LB\_Partition geometric partitioner interface in Zoltan e.g. HSFC}
	\ElsIf  { ( partitioningType == graph-based ) }
	\State  myLeafConnectivity, weights = \texttt{getConnectivityAndWeights}( myProcessLeafIds, mesh )
	\State myProcessLeafIds = \texttt{improveLoadBalance}( myLeafConnectivity, weights )
	\State \Comment{call to LB\_Partition graph-based partitioner interface in Zoltan e.g PHG}
        \EndIf
	\State \textbf{end}
    \State \textbf{return} myProcessLeafIds
    \EndFunction
\end{algorithmic}
\end{algorithm}

As previously mentioned, the \textit{LeafElementPartitioner} is responsible for determining the weight, computational cost,  of individual leaf elements. This is of particular importance in the context of multi-level $hp$-refinement, since the number of basis functions supported within a leaf element varies greatly depending on the number of overlay meshes, as shown in Figure \ref{fig::multiLevelFCM::integration}. Moreover, the number of integration points within a leaf element differ when composed integration in the finite cell method is used. This difficulty is overcome by considering the time complexity of the matrix-matrix multiplications needed to compute the stiffness matrix of a single leaf element, since this operation is the main bottleneck during computations. The integration time and consequently weight $w$ of a leaf element is directly proportional to the number of quadrature points $n_{GP}$ and the third power of the number of active shape functions denoted by N. The cubic time complexity in terms of $N$ can be illustrated by the fact that $\mathcal{O}(N^3)$ multiplications have to be performed to compute the bilinear term in Equation \ref{eq::multiLevelFCM::integration}. The weight of a leaf element is thus approximated as $w \approx  n_{GP} \cdot N^{3}$ in lines 6 and 10 of Algorithm \ref{alg:distribution_of_integration_domains}, as $N$ and $n_{GP}$ are known before integration. In concrete simulations $w$ is normalized with the weight of an unrefined element yielding a modified weight $w^* = w / w_0$. This modification aims to factor in influences not taken into account during the derivation of $w$. 
 \subsubsection{Integration of the local linear system}
 The integration of the local linear system is performed in a loop over the local set of active elements $\Omega^{h}_{proc}$. A second level of parallelism is made available through an OpenMP loop over the integration points within a leaf element. Each process assembles the contributions of its leaf elements into an intermediate linear system. The intermediate stiffness matrix for example, can be efficiently assembled using a local sparsity pattern based solely on the degrees of freedom present in $\Omega^{h}_{proc}$. Boundary conditions are then applied independently on the intermediate linear system by each process.  
\subsubsection{Distribution of degrees of freedom among processes}
Before the global linear system can be assembled, the question of degree of freedom ownership has to be addressed. In this step, the degrees of freedom have to be uniquely distributed among the processes. To this end, two distribution algorithms are proposed. Degrees of freedom can either be distributed contiguously, where each process is assigned a set of consecutive unknowns, or be partitioned using a graph-based approach, which aims to minimize the amount of data communicated during distributed assembly. The first algorithm is easy to implement but could result in increased communication costs during distributed assembly, as the current distribution of leaf elements is not taken into account. The results presented in Section \ref{numericalResults}, however, demonstrate that this additional overhead has no major relevance. This may be attributed to the use of a rather moderate number of MPI processes ($<1000$). The graph based approach takes the current distribution of leaf elements into account in order to minimize the amount of information transfered during distributed assembly. This process is summarized in Algorithm \ref{alg:distribution_of_degrees_of_freedom} where a graph containing the relations between degrees of freedom and leaf elements is used to assign a degree of freedom to the process that contains the highest number of leaf elements associated with that particular degree of freedom. This in simple terms translates into the following principle: The process that did the most work for a certain degree of freedom, gets to keep it. In the case that two or more processes have an equal number of integration domains corresponding to a certain degree of freedom, the process with a lower MPI-rank is assigned the degree of freedom.  
\subsubsection{Assembly of the global linear system}\label{assembleDistributedLS}
As previously mentioned, the proposed parallelization strategy avoids redundant computations associated with ghost elements. This results in non-local entries in a process' intermediate linear system, which need to be communicated to other processes, \textit{cf.} Figure \ref{fig::parallelImplementation::comparisonDomainDecomposition}b. In order to perform a distributed assembly in an efficient way, the algorithms and data structures available in the distributed linear algebra packages in \texttt{Trilinos} \cite{Trilinos} or \texttt{Hypre} \cite{Falgout02hypre:a} are used. These packages allow non-local matrix and vector entries to be communicated in an efficient way and are widely used in different $hp$-codes \cite{BangerthHartmannKanschat2007,mfem-library}.   
\subsubsection{Solving the global linear system}
Once assembled, the distributed linear system can be solved using different parallel solvers. The implementation provides an interface to parallel direct solvers available in \texttt{Trilinos} and parallel iterative solvers and preconditioners in \texttt{Hypre} and \texttt{Trilinos}. The choice of a suitable solver is dependent on different factors such as the problem type e.g. linear elasticity or Poisson, spatial dimension, size and the conditioning of the stiffness matrix.   

\subsubsection{Distributed post-processing}
Post-processing is carried out independently on the local partition $\Omega^{h}_{proc}$ of each process. We use the parallel hierarchical data format (PHDF5) \cite{hdf5} and MPI-I/O to perform distributed I/O so as to simplify the management of the large datasets. We adopt a strategy where all processes write their result data into a single HDF5-file while a single process is responsible for writing an XML-file with the corresponding meta-data. This approach reduces the pressure on the filesystem and allows simple visualization as huge datasets in separate files do not have to be combined for visualization.   
\begin{algorithm}[H]
	\caption{ Distribution of degrees of freedom among processes }
\label{alg:distribution_of_degrees_of_freedom}
\begin{algorithmic}[1]
    \Function{\texttt{distribute\_degrees\_of\_freedom} }{ mesh, myProcessLeafIds }
    \If { ( dofDistribution == contiguous ) }
    \State myProcessDofs = \texttt{assignContiguousDofsToProcess}( mesh )
    \ElsIf  { ( dofDistribution == graph-based ) }
    \State  leafDofMap = \texttt{getDofLeafElementConnectivity}( mesh, myProcessLeafIds )
    \State  nonLocalLeafIds = \texttt{getNonLocalLeafElementsWithSharedDofs}( leafDofMap )
    \State  nonLocalLeafProcessIds = \texttt{getProcessIdOfLeafElement}( nonLocalLeafIds ) 
    \State  \textbf{for} ( iDof : mesh.activeDegreesOfFreedom ) \Comment{Decide which dofs should be kept}
    \State  \hspace{3mm} \textbf{if} ( \texttt{queryDegreeOfFreedom}( iDof, leafDofMap, nonLocalLeafProcessIds ) )
    \State  \hspace{8mm} myProcessDofs.insert( iDof )
    \State  \hspace{3mm} \textbf{end}
    \State  \textbf{end}
        \EndIf
    \State \textbf{end}
    \State \textbf{return} myProcessDofs
    \EndFunction
\end{algorithmic}
\end{algorithm} 

\begin{algorithm}[h!]
	\caption{ Summary of the simulation pipeline }
\label{alg:simulation_pipeline}
\begin{algorithmic}[1]
    \Function{\texttt{solve} }{ }
    \State mesh = \texttt{createMesh}()
    \State \textbf{for} ( iStep : timeSteps )
    \State \hspace{3mm} allLeafElements = \texttt{refineElements}( mesh.baseElements ) 
    \State \hspace{3mm} myProcessLeafIds = \texttt{getActiveLeafElementsOnProcess}( mesh.baseElements ) \Comment{Alg. \ref{alg:distribution_of_integration_domains}}
    \State \hspace{3mm} \texttt{registerActiveLeafElementsOnProcess}( myProcessLeafIds ) 
    \State \hspace{3mm} \textbf{for} ( iLeafElement : myProcessLeafElements )
    \Comment{Integration loop over active leaf elements}
    \State \hspace{8mm} \textbf{\#pragma omp for}  
    \State \hspace{8mm} \textbf{for} ( intPoint : Integrationpoints )
    \State \hspace{12mm} matrices =+ \texttt{calculateLeafElementMatrices}()
    \Comment{e.g stiffness, mass matrix}
    \State \hspace{12mm} rhs =+ \texttt{calculateRightHandSide}() 
    \State \hspace{8mm}\textbf{end}
    \State \hspace{8mm} intermediateLinearSystem = \texttt{scatterIntoIntermediateLinearSystem}( matrices, rhs ) 
    \State \hspace{3mm} \textbf{end}
    \State \hspace{3mm} myProcessDofs = \texttt{distributeDegreesOfFreedom}( mesh, myProcessLeafIds ) \Comment{Alg. \ref{alg:distribution_of_degrees_of_freedom}}
    \State \hspace{3mm} globalLinearSystem  = \texttt{initializeGlobalLinearSystem}( myProcesDofs ) 
    \State \hspace{3mm} globalLinearSystem.\texttt{assembleDistributedLinearSystem}( intermediateLinearSystem ) 
    \State \hspace{3mm} \texttt{solveDistributedLinearSystem}( globalLinearSystem ) 
    \State \hspace{3mm} \texttt{postProcessResults}() 
    \State \textbf{end} 
    \State \hspace{-5mm}\textbf{return}
    \EndFunction
\end{algorithmic}
\end{algorithm} 

\section{Numerical Results}\label{numericalResults}

\renewcommand{\rootDir}{numericalResults}
\renewcommand{\graphDir}{\rootDir/graphs}
\renewcommand{\dataDir}{\rootDir/data}
\renewcommand{\picsDir}{\rootDir/pics}

This section highlights various aspects of the proposed parallel scheme. Starting with a simple example in two dimensions, we investigate the performance of the presented algorithms focusing on their scalability and execution time for different problem sizes. Next, the performance of the scheme is shown in a complex three dimensional transient example involving complicated refinement patterns.  Further, an example involving FCM on a domain of complex shape is considered, to show the scheme's applicability to problems of engineering relevance. Although our implementation can be run within a hybrid framework as portrayed in Algorithm \ref{alg:simulation_pipeline}, we restrict our numerical examples to the MPI-flat performance of the code. 

The examples are computed on the CoolMAC cluster at Technical University of Munich. Each node has a dual socket Intel Sandy Bridge-EP Xeon E5-2670 architecture with a total of 16 processors and 128GB memory per node. QDR infiniband connects the nodes to each other. Moreover, the following library and compiler versions are used: Trilinos 12.6.1, Hypre 2.11.1, Zoltan v3.83, Boost 1.56 and IntelMPI compiler version 5.0 alongside the gcc compiler version 4.9 with the compiler flags -O3 and -funroll-loops. Logging of the execution time is performed using the cpu\_timer implementation in the Boost library \cite{Schling:2011:BCL:2049814}, while the memory consumption during runtime is obtained directly from the operating system via a call to the process status. Although our scheme offers an interface to different parallel direct and iterative solvers, only one solver is used for all numerical examples for the sake of brevity. All examples are solved using the parallel conjugate gradient solver with a multi-grid preconditioner available in \texttt{Hypre} \cite{Falgout02hypre:a}.

\subsection{2D singular benchmark}
\begin{figure}[!htb]
  \begin{minipage}{0.35\textwidth}
  \begin{center}	
      \includegraphics[width=0.9\textwidth]{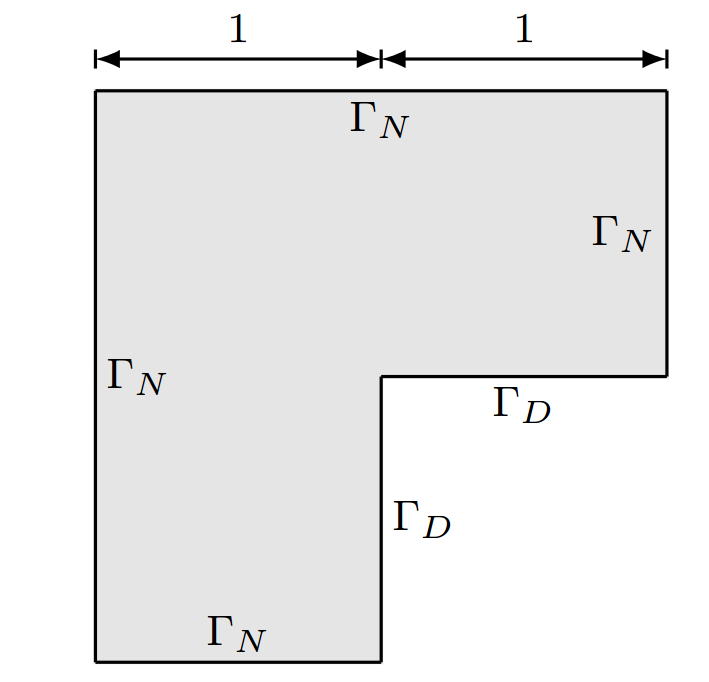}%
  \end{center}
  \end{minipage}
  \begin{minipage}{0.6\textwidth}
  \begin{center}	
\begin{align*}
	\phi &= 0  &\forall \ (r,\theta) \in \Gamma_D \, , \nonumber \\
\nabla \phi &= \frac{2}{3}r^{-\frac{4}{3}} \begin{bmatrix} x \sin(\frac{2}{3}\theta) - y \cos(\frac{2}{3}\theta) \\ y \sin(\frac{2}{3}\theta) + x \cos(\frac{2}{3}\theta)  \end{bmatrix} &\forall \ (r,\theta) \in \Gamma_N \, ,
\end{align*}
\begin{equation*}
	\phi(r,\theta) = r^{\lambda} \sin(\frac{2}{3}\theta) \ \ \  \forall \ (r,\theta) \in \Omega \ \ \ \text{with} \ \ \ \lambda = \dfrac{2}{3} \, .
\end{equation*}
  \end{center}	
  \end{minipage}
   \caption{Problem setup of the L-shaped domain}
   \label{fig::numericalResults::LshapedDomain}
\end{figure}

The L-shaped domain benchmark considered by Szab\'{o} and Babu\v{s}ka in \cite{Szabo1991} is taken as the first example. It aims to solve the Laplace problem $\Delta \phi = 0$ on the domain depicted in Figure \ref{fig::numericalResults::LshapedDomain}, subjected to the boundary conditions as shown. An analytical solution $\phi(r,\theta)$ can be derived for this problem with the origin at the re-entrant corner, see e.g \cite{Demkowicz2007}, and is also given in Figure \ref{fig::numericalResults::LshapedDomain}. 

\subsubsection{Scalability of the simulation pipeline}\label{RunAandRunB}
To investigate the scalability of the parallel implementation for different problem sizes, computations on two fixed discretizations of the L-shaped domain are considered. Firstly, each of the three quadrants of the L-shaped domain is discretized uniformly with 16 by 16 elements resulting in a coarse base mesh comprising of 768 elements. This mesh is refined recursively in 5 steps towards the domain's re-entrant corner and studied. This test case is termed as \textit{run A} and consists of 813 leaf elements. A polynomial order of $p=18$ is chosen, ensuring that the measured times are in the range of seconds. The second test case, \textit{run B}, consists of a finer mesh with 49\,152 base elements, obtained by discretizing each quadrant with 128 by 128 elements. This mesh is also refined in 5 steps yielding a total of 49\,197 leaf elements. A polynomial degree $p=10$ is chosen in this test case. The computational domains are partitioned among different numbers of processes and the execution time for various components in the simulation pipeline is monitored. 

% scaling curves
% \begin{figure}[H]
%	   \begin{center}
%	         %\tikzset{external/remake next}%
%	         \nextfigurename{fcm}%
%	       \includegraphics[width=0.5\textwidth]
%               {\graphDir/scalabilityLshapedDomain.tikz}%
%           \end{center}
%   \caption{Problem setup}
%    \label{fig::parallelImplementation::AdhoC++2}
%\end{figure}

%%\begin{figure*}[h]
%%  \begin{center}
%
%I want to show the scalability of pure MPI computations and hybrid computations
%for different problem sizes. 
%
%
%
\begin{figure*}[h]
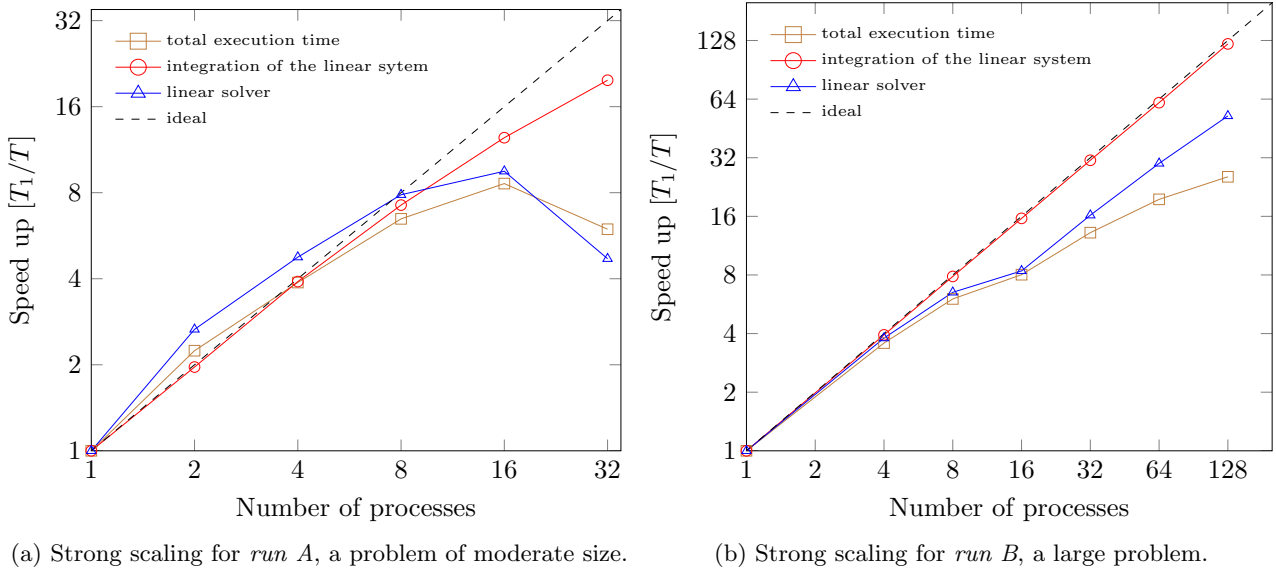

  \begin{center}
	  \subfloat[Strong scaling for \textit{run A}, a problem of moderate size.]
  	{%
	  %\tikzset{external/remake next}%
      \includegraphics[width=\halfTextWidth,
        height = 7cm]%  
        {\graphDir/scalabilityLshapedDomainSmall.tikz}%
	\label{fig::numericalResults::speedUpRuna}
    }%
    \subfloat[Strong scaling for \textit{run B}, a large problem.]
  	{%
	  %\tikzset{external/remake next}%
     \includegraphics[width=\halfTextWidth,height = 7cm]{\graphDir/scalabilityLshapedDomain.tikz}%
     \label{fig::numericalResults::speedUpRunb}
    }%
    \hfill%   
  \end{center} 
  \caption{Strong scaling of the parallel multi-level $hp$-refinement for two test cases.}
  \label{fig::numericalResults::strongScalingLSD}
\end{figure*} 

Figure \ref{fig::numericalResults::strongScalingLSD} shows the strong scaling of the parallel implementation for the test cases \textit{run A} and \textit{run B}. The most time consuming components in the simulation pipeline, integration of the stiffness matrices and solving of the global system, together with the total execution time are considered. In Figure \ref{fig::numericalResults::speedUpRuna} the speed up in $\textit{run A}$ for the mentioned components is shown, with each component scaling almost linearly or better up to four processes. In this setup with four processes, each process integrates around 200 leaf elements. The speed up of the integration time is still relatively good at a process count of 8, with as little as 100 leaf elements per process, but decreases gradually as the number of processes and consequently the number of integration domains per process decreases.  The linear solver scales well up to 8 processes, but suffers a reduction in parallel efficiency due to limited computational work and added communication costs when the number of processes is increased. The scalability of the whole simulation is largely influenced by the scalability of the solver, as shown by the strong correlation of the two curves. These curves are, however, not identical due to other components in the simulation pipeline such as the serial parts of the code. It should be noted that the assembly of the distributed linear system is not included in the curves in Figure \ref{fig::numericalResults::strongScalingLSD}.

Next, we consider the performance of the parallel implementation for large problems in \textit{run B}. The setup comprises of 5 million degrees of freedom with the process count varying from 1 to 128. The increased number of integration domains per process has a positive effect on the integration algorithm, as near to perfect scalability is achieved up to a process count of 128, see Figure \ref{fig::numericalResults::speedUpRunb}. This number is still significantly lower than that of some conventional approaches which need more elements per process, up to $10^5$, to hide the overhead introduced by redundant computations \cite{BangerthHartmannKanschat2007}. The $\texttt{Hypre}$ solver scales well to 8 processes, but suffers a loss in efficiency with 16 processes. Its parallel performance however improves when 32 processes are used. This is attributed to cache effects as the problem is now computed on two nodes. The overall scalability of the problem is closely coupled with the scalability of the linear solver similar to the results from \textit{run A}.   

\begin{figure*}[h]
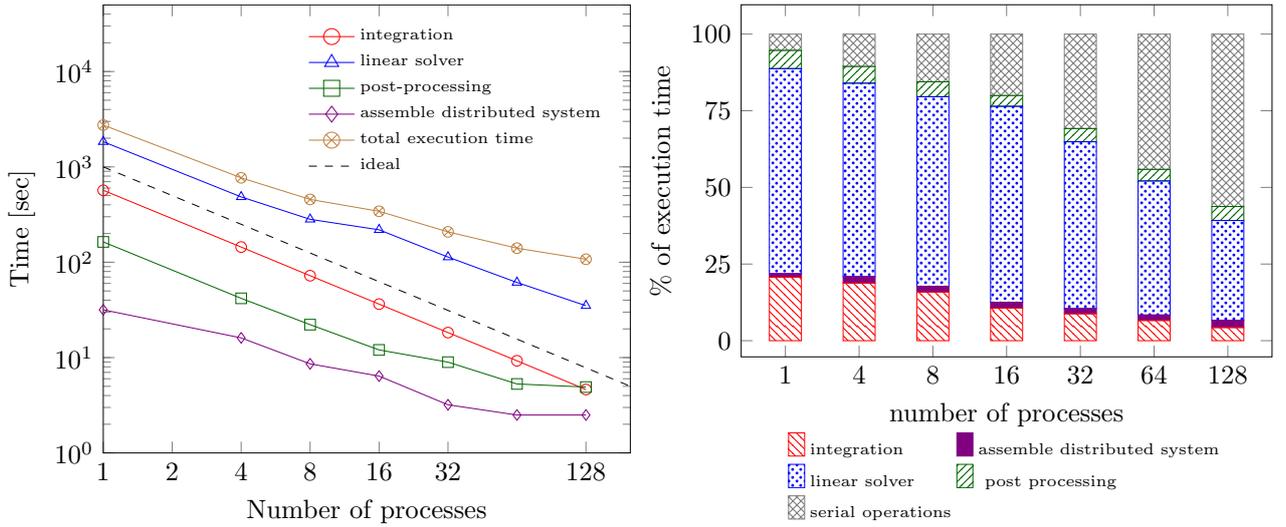

  \begin{center}
	  \subfloat[Execution time of different components in \textit{run B}.]
  	{%
	  %\tikzset{external/remake next}%
      \includegraphics[width=\halfTextWidth,
        height = 7cm]%  
        {\graphDir/timingLshapedDomain.tikz}%
	\label{fig::numericalResults::timingLSDRunb}
    }%
    \subfloat[Contribution of the different components to the total execution time in \textit{run B}.]
  	{%
	  %\tikzset{external/remake next}%
     \includegraphics[width=\halfTextWidth,
        height = 7cm]% 
        {\graphDir/percentageLshapedDomain.tikz}%
	  \label{fig::numericalResults::percentageLSDRunb}
    }%
    \hfill%   
  \end{center} 
  \caption{Runtime analysis of the simulation pipeline for \textit{run B}.}
  \label{fig::numericalResults::runTimeAnalysisRunb}
\end{figure*} 

A more detailed analysis of the simulation pipeline is performed using \textit{run B} as a basis. Figure \ref{fig::numericalResults::timingLSDRunb} shows the execution times of different components plotted against the number of processes used, while Figure \ref{fig::numericalResults::percentageLSDRunb} shows their contribution to the total execution time. These diagrams show that the communication of the non-local entries of the stiffness matrices to other processes during the assembly of the distributed system, an integral part of the parallelization scheme (see Section \ref{assembleDistributedLS}), does not significantly influence the simulation's overall scalability and constitutes at most 5\% of the execution time. Its contribution to the overall time is, however, expected to increase when the number of processes increases above 128. This would carry the same computational cost as the integration of the stiffness matrices. Furthermore, the serial part of the code would dominate the total computation time. The problem size is in this case too small to effectively utilize the available resources.      

\subsubsection{Influence of the polynomial order on the integration algorithm}
The results from \textit{run A} and \textit{run B} show almost perfect scalability in the integration of the stiffness matrices. Both cases, however, utilize a rather high polynomial order so as to increase the amount of computational work per process. The behavior of the algorithm for low and moderate values of $p$, however, remains to be shown. We investigate this by computing the example at hand with different polynomial orders where  $p \in \lbrace 1,\dots ,7 \rbrace$. By keeping the number of processes constant at 8 and increasing the number of mesh elements, we are able to monitor the development of the parallel efficiency of the integration algorithm for different polynomial orders and elements numbers.
\begin{figure}[H]
	   \begin{center}
	         %\tikzset{external/remake next}%
	         %\nextfigurename{fcm}%
	       \includegraphics[width=0.5\textwidth, height= 7cm]
               {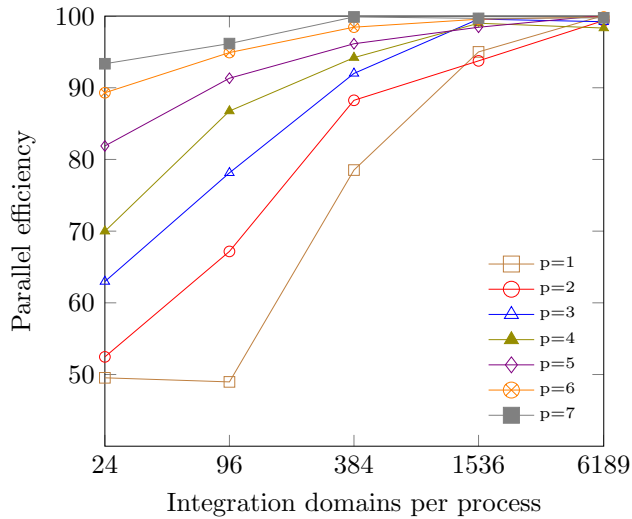}%
           \end{center}
   \caption{Parallel efficiency of the integration algorithm for different polynomial orders.}
    \label{fig::parallelImplementation::pstudyLshapedDomain}
\end{figure}
The results of the study are presented in Figure \ref{fig::parallelImplementation::pstudyLshapedDomain}. The parallel efficiency of the integration algorithm increases when $p$ is raised, due to the increase in computational work, and can be further improved by increasing the number of elements per process. This integration scheme is therefore also suitable for problems with low or moderate polynomial orders.    
\subsubsection{Memory requirements}
Another important aspect of any parallel scheme is the amount of random-access memory (RAM) needed during a computation. Using \textit{run B} from Section \ref{RunAandRunB} as a basis, we analyze the peak total memory needed during a simulation as well as the average and maximum memory usage of a single process. These values are not constant but only represent the highest memory requirements at a single juncture in the simulation. Peak memory consumption is in general problem dependent, and is reached for the problem at hand during the solution of the linear system. 
\begin{figure}[H]
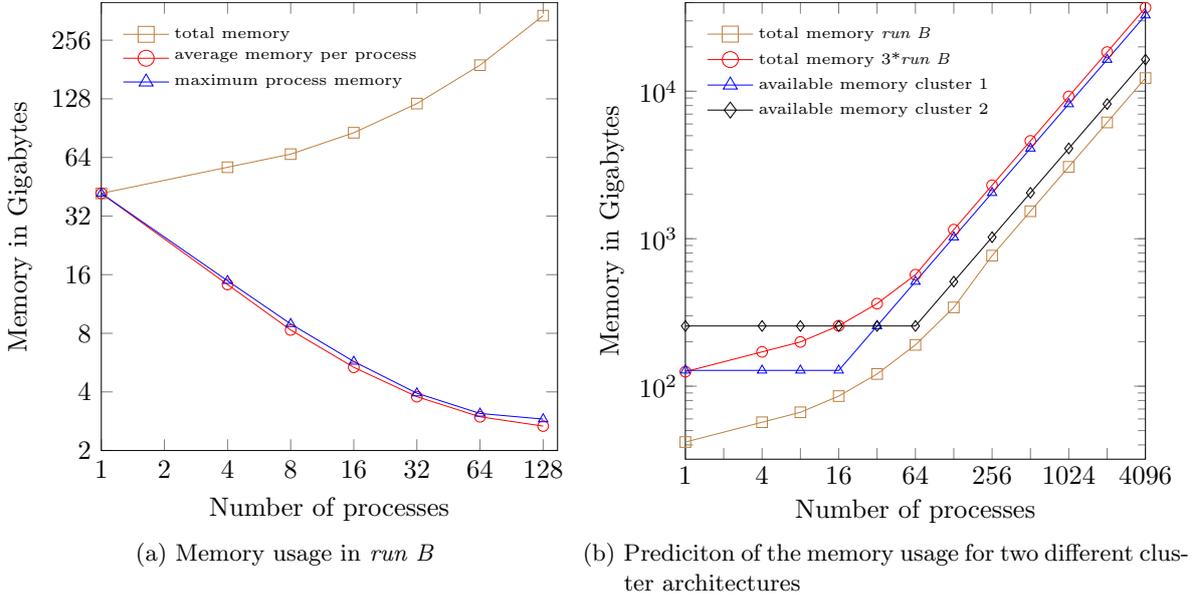

   \begin{center}
        %\tikzset{external/remake next}%
        %\nextfigurename{fcm}%
	   \subfloat[Memory usage in \textit{run B}]
     {
       \includegraphics[width=0.45\textwidth, height = 7cm]{\graphDir/memoryLSD.tikz}%
       \label{fig::numericalResults::memoryLshapedDomainRunB}
     }%
     \subfloat[Prediciton of the memory usage for two different cluster architectures]
     {
       \includegraphics[width=0.46\textwidth, height = 7cm]{\graphDir/memoryLSD-prediction.tikz}%
       \label{fig::numericalResults::memoryLshapedDomainPrediction}
     }%
    \hfill%
    \end{center}
    \caption{ Analysis of the memory requirements of the parallel multi-level scheme.}
     \label{fig::numericalResults::memoryLshapedDomain}
\end{figure}

The memory needed by a single process decreases with an increase in the number of processes as shown in Figure \ref{fig::numericalResults::memoryLshapedDomainRunB}, this reduction rate is, however, sub-linear and leads to an overall increase in the total memory needed. The minimal difference between the average and maximum process memory usage reveals that the proposed scheme is also able to achieve good balance in terms of memory consumption. Nevertheless, the memory needed by a single process starts to level off at a process count of 64. We attribute this behavior to the shared mesh, which is duplicated on every process and thus a natural lower bound. This leveling off in turn, would lead to a linear increase in the total memory at a process count beyond 128. 

Following the above observation, the total memory required for simulations involving 128 to 4096 processes is approximated under the assumption of a limit shared mesh size of 3GB . The total memory usage in Figure \ref{fig::numericalResults::memoryLshapedDomainRunB} is extrapolated and compared in Figure \ref{fig::numericalResults::memoryLshapedDomainPrediction} to the amount of memory available on two different cluster architectures, cluster 1 with nodes comprising 16 Intel SandyBridge processors and 128GB of memory and cluster 2 with nodes made up of 64 AMD Bulldozer Opteron processors and 256GB of memory. Figure \ref{fig::numericalResults::memoryLshapedDomainPrediction} shows that \textit{run B} could be successfully run on both clusters with between 128-4096 processes without exceeding the available memory resources. This, however, does not hold when the setup size is tripled, this corresponds to a shared mesh of 9 GB and a total of 17 million degrees of freedom if the polynomial degree $p=10$ is assumed constant.     

\subsection{3D benchmark}

\begin{figure}[!htb]
  \begin{minipage}{0.4\textwidth}
   \centering	  
   \begin{center}
      \includegraphics[width=0.9\textwidth]{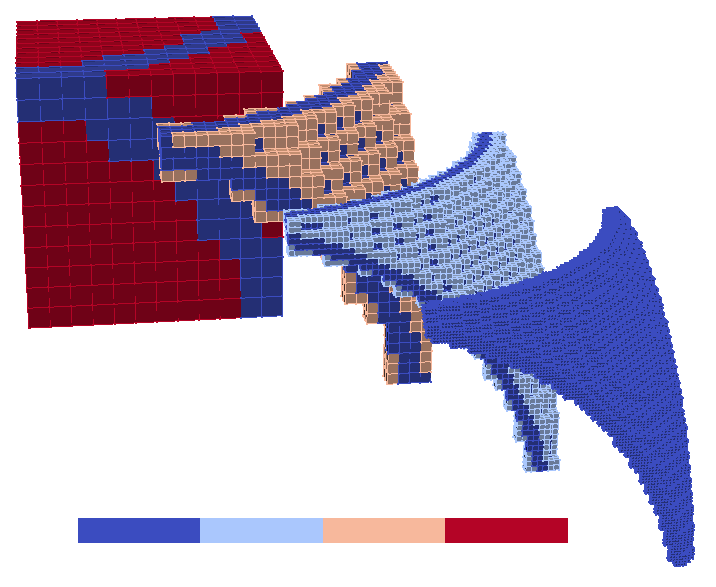}%
      \put(-119,-3){\small Ansatz order}
      \put( -51,19){\small 8}
      \put( -80,19){\small 6}
      \put(-109,19){\small 4}
      \put(-140,19){\small 1}
    \end{center}	 
    (g) Graded mesh at $t$=5.
  \end{minipage}
  \begin{minipage}{0.58\textwidth}
    \begin{minipage}[t]{0.4\textheight}	  
     \centering
     \begin{center}
       \subfloat[$t=2$]{ \includegraphics[width=0.27\textwidth]{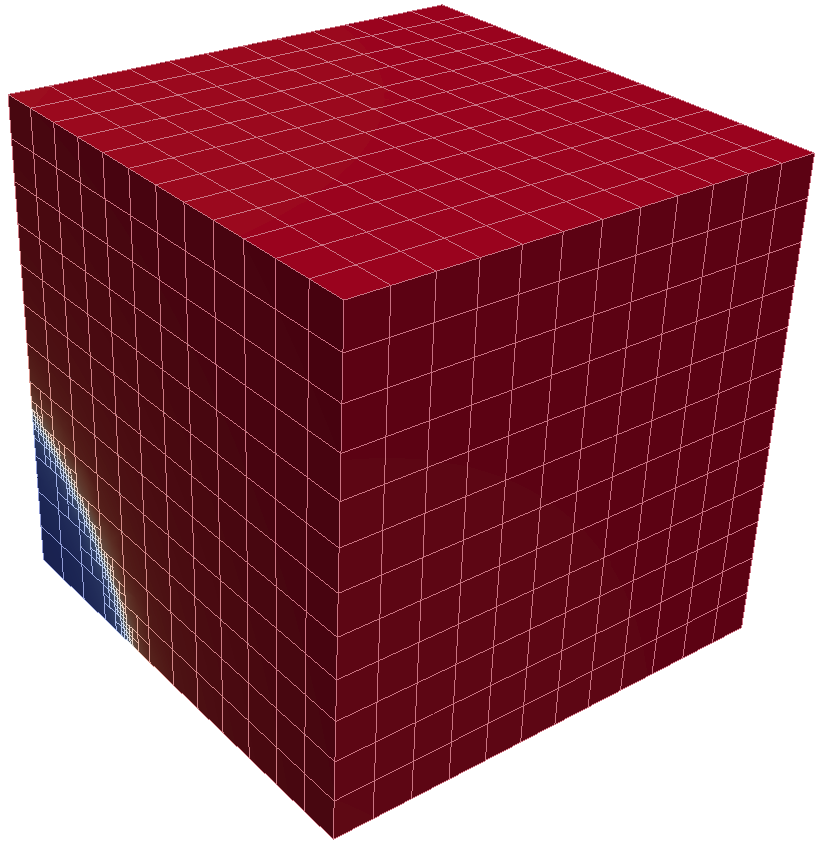} }%
       \subfloat[$t=4$]{ \includegraphics[width=0.27\textwidth]{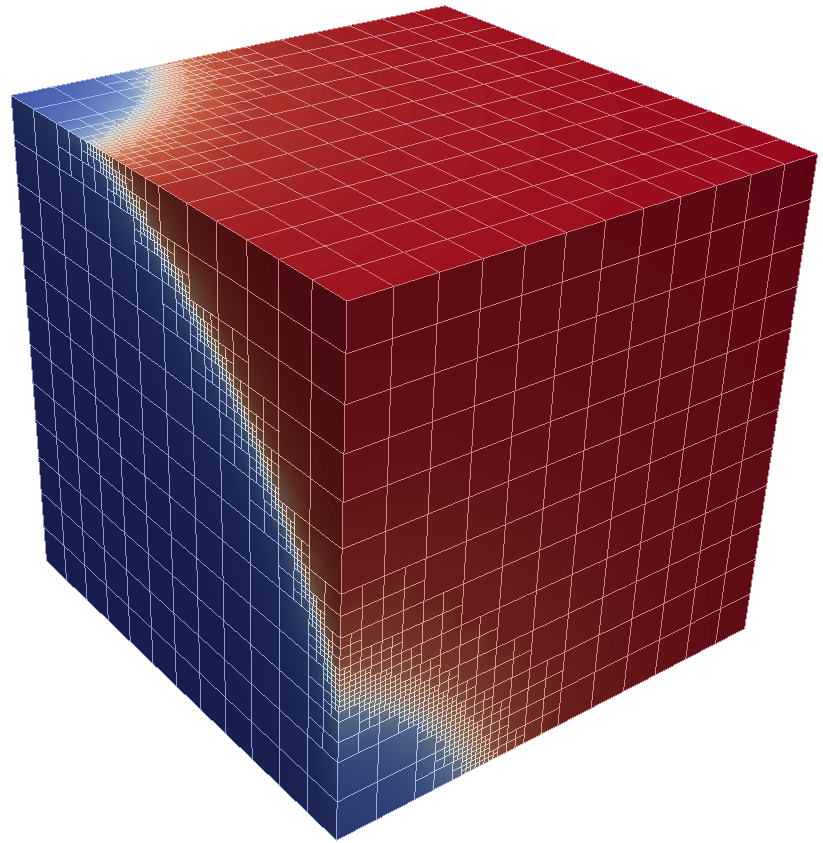} }%
       \subfloat[$t=5$]{ \includegraphics[width=0.27\textwidth]{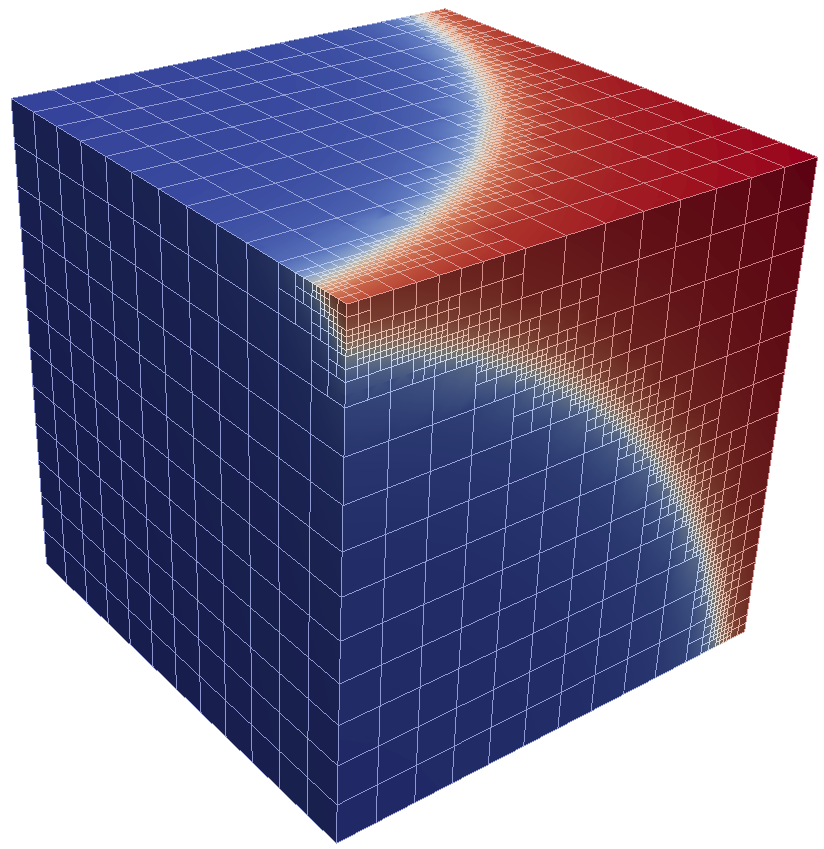} }% 
     \end{center}	
       Solution of the shock problem.
    \end{minipage}	  
    \begin{minipage}[t]{0.4\textheight}
     \centering	    
     \begin{center}
       \subfloat[$t=2$]{ \includegraphics[width=0.27\textwidth]{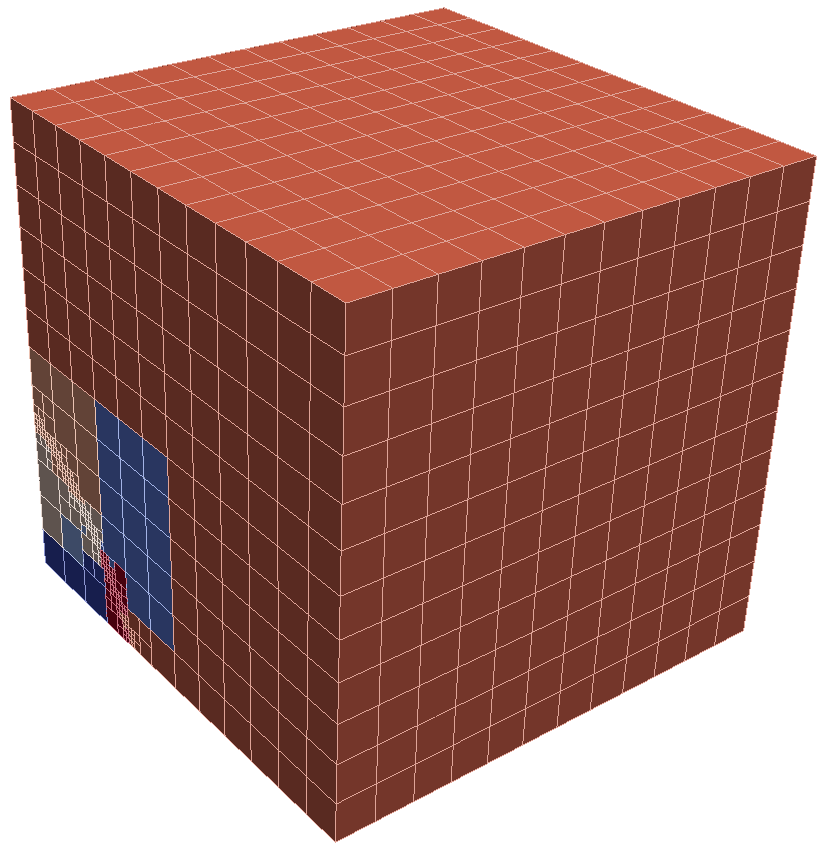} }%
       \subfloat[$t=4$]{ \includegraphics[width=0.27\textwidth]{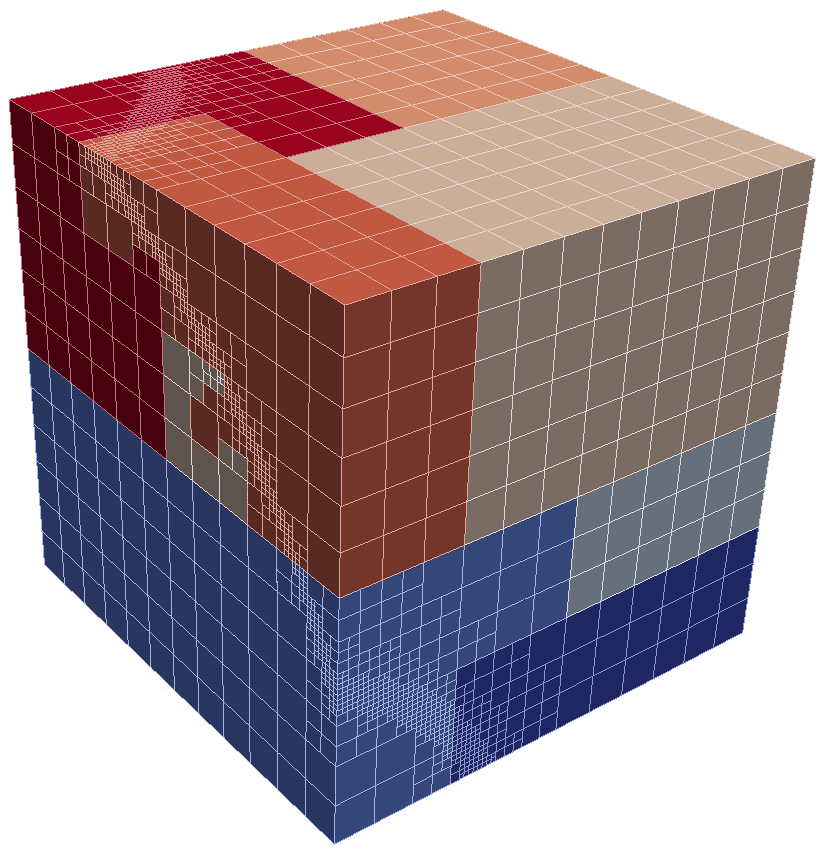} }%
       \subfloat[$t=5$]{ \includegraphics[width=0.27\textwidth]{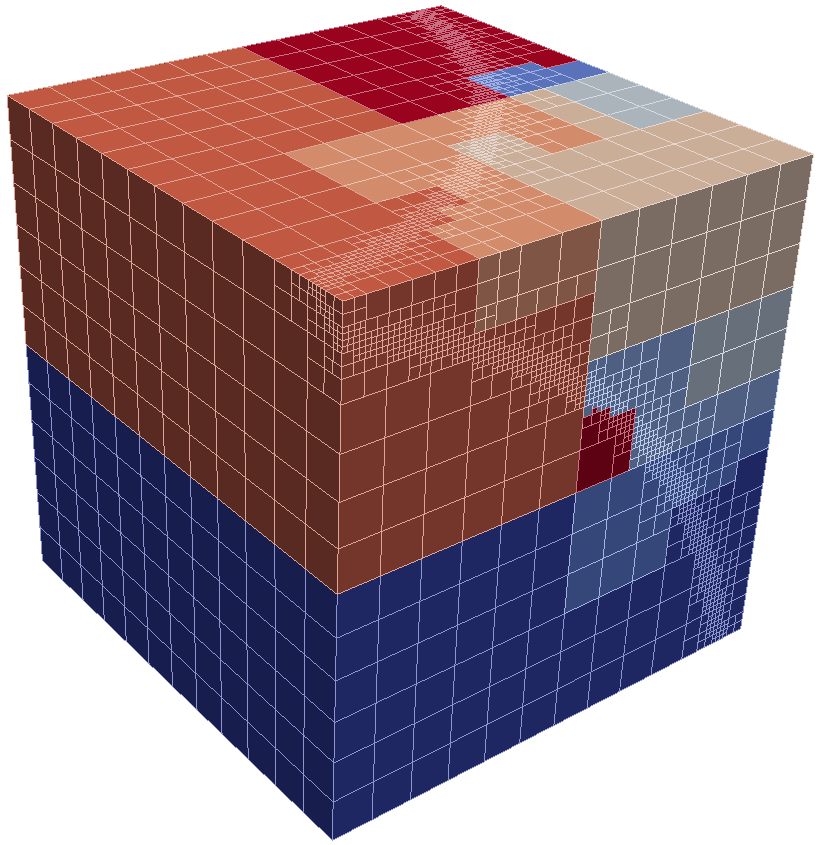} }%
     \end{center}	
     Distribution of the domain among 8 processes.
   \end{minipage}	  
  \end{minipage}
  \caption{Problem setup of the modified shock problem}
   \label{fig::numericalResults::shockProblem}
\end{figure}

The applicability of the parallel scheme to three dimensional problems is studied in this example. To this end, the shock problem considered in \cite{Rachowicz2006,Zander2016} is chosen due to the complex mesh refinement patterns involved. It entails solving the Poisson equation on a unit cube $\Omega = [0,1]^3$ subjected to Dirichlet and Neumann boundary conditions derived from the manufactured solution
\begin{equation}
	u = \tan^{-1} ( \alpha ( r - r_0 )  ) \ \ \text{with} \ \ r_0 = \sqrt{3} \ \ \text{and} \ \ \alpha \in \lbrace 40, 80, 160 \rbrace \, .
	\label{eq::u}
\end{equation}
$u$ is specially chosen to represent a shock-like function as illustrated in Figure \ref{fig::numericalResults::shockProblem}. The term $r$ is a radial coordinate about a shifted origin given by the equation,
\begin{equation}
	r = \sqrt{(x-x_0)^2 + (y-y_0)^2 + (z-z_0)^2 } \ \ \text{with} \ \ x_0 = y_0 = z_0 = -0.25 \, ,
	\label{eq::r}
\end{equation}
while $\alpha$ is a factor determining the sharpness of the shock. The interested reader is referred to \cite{Zander2016} for a detailed derivation of the applied boundary conditions and multi-level $hp$-convergence properties. In this numerical example, the original shock problem is modified to allow for a time dependent initial radius $r_0=r_0(t)$ in Equation \ref{eq::u}.  The performance of our scheme in a transient setup can thus be investigated, for a sharpness $\alpha=160$ and different values of $r$ as illustrated in Figure \ref{fig::numericalResults::shockProblem}a-c.
%
%\begin{figure}[H]
%      \begin{center}
%                %\tikzset{external/remake next}%
%                %\nextfigurename{fcm}%
%              \includegraphics[width=0.3\textwidth]
%                  {\picsDir/solution-shock.png}%
%          \end{center}
%      \caption{Prescribed solution: shock problem.}
%     \label{fig::numericalResults::shockProblem}
%  \end{figure}
 
\begin{figure}[H]
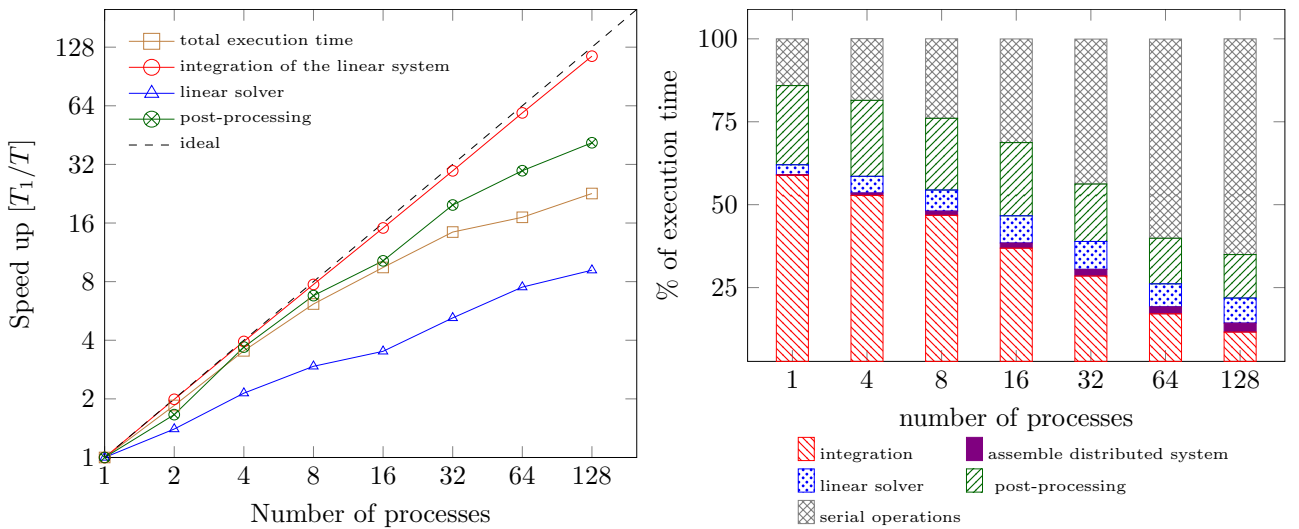

  \begin{center}
   \subfloat[Strong scaling of the total time in selected routines.]{%
     %\tikzset{external/remake next}%
     \includetikz[width=0.5\textwidth, height = 7cm]{\graphDir/scalabilityShockTransient}%
     \label{fig::numericalResults::speedUpShock}
    }%
   \subfloat[Contribution of different components to the total execution time.]{%
     %\tikzset{external/remake next}%
     \includetikz[width=0.5\textwidth, height = 7cm]{\graphDir/percentageShockTransient}%
     \label{fig::numericalResults::percentShock}
    }%
    \hfill%   
  \end{center} 
  \caption{Analysis of the overall performance in the shock problem.}
  \label{fig::numericalResults::shockProblemPerformance}
\end{figure}

The example at hand beings in the first time step with an initial mesh of $12^3$ base elements and a uniform polynomial degree $p=8$. Moreover, the trunk space following \cite{Szabo1991} is used. This computational mesh is refined in every time step towards the shock with a refinement depth of three as illustrated in Figure \ref{fig::numericalResults::shockProblem}d-f and a total of five time steps are computed. A graded polynomial order is applied to the elements, \textit{cf.}\,\cite{Zander2016}, as illustrated in Figure \ref{fig::numericalResults::shockProblem}g for the final time step, where the polynomial order is decreased when increasing the refinement level. A graded mesh posses a challenge for any scheme in terms of load balancing, as the number of basis functions within leaf elements vary greatly. The number of leaf elements and unknowns within each time step are summarized in Table \ref{tab::numericalResults::shockProblemSetup}. The performance of the simulation pipeline is monitored in every time step as in the previous example.
\begin{table}[H]
 \centering
 \begin{tabular}{cccccc}
  \toprule
   Time step      & 1 & 2 & 3 & 4 & 5 \\ \midrule
  $r_0$          & $0.2\sqrt{3}$ & $0.4\sqrt{3}$ & $0.6\sqrt{3}$ & $0.8\sqrt{3}$ & $\sqrt{3}$ \\ 
   No. degrees of freedom  & 158437 & 191329 & 240028 & 266506 & 214576 \\
   No. leaf elements & 6474 & 19095 & 38604 & 48390 & 27201 \\
 \bottomrule
 \end{tabular}
 \caption{Time step information in the modified shock problem}
 \label{tab::numericalResults::shockProblemSetup}
 \end{table}
 The results of the present study are shown in Figure \ref{fig::numericalResults::shockProblemPerformance}. The total time spent in various routines was monitored over the 5 time steps and used as a basis for computing the overall scalability and the contribution of individual components to the total execution time. A similar pattern in the scalability and simulation pipeline breakdown is seen as in the two dimensional case. Although the mesh in each time step consists of elements with greatly varying polynomial orders, our partitioning algorithm is able to achieve good balance and allows scalability of the integration algorithm up to about 128 processes. The linear solver does not scale well in this simulation. This is due to a rather small problem size of about 200\,000 degrees of freedom. The influence of the serial components of the code also increases with an increase in the number of MPI processes as shown in Figure \ref{fig::numericalResults::percentShock}.  

\subsection{Parallelization of the multi-level $hp$-FCM}

\begin{figure}[H]
 \centering
          \subfloat[\footnotesize Loading of the bone.]
        {%
          %\tikzset{external/remake next}%
      \includegraphics[width=0.25\textwidth]%  
        {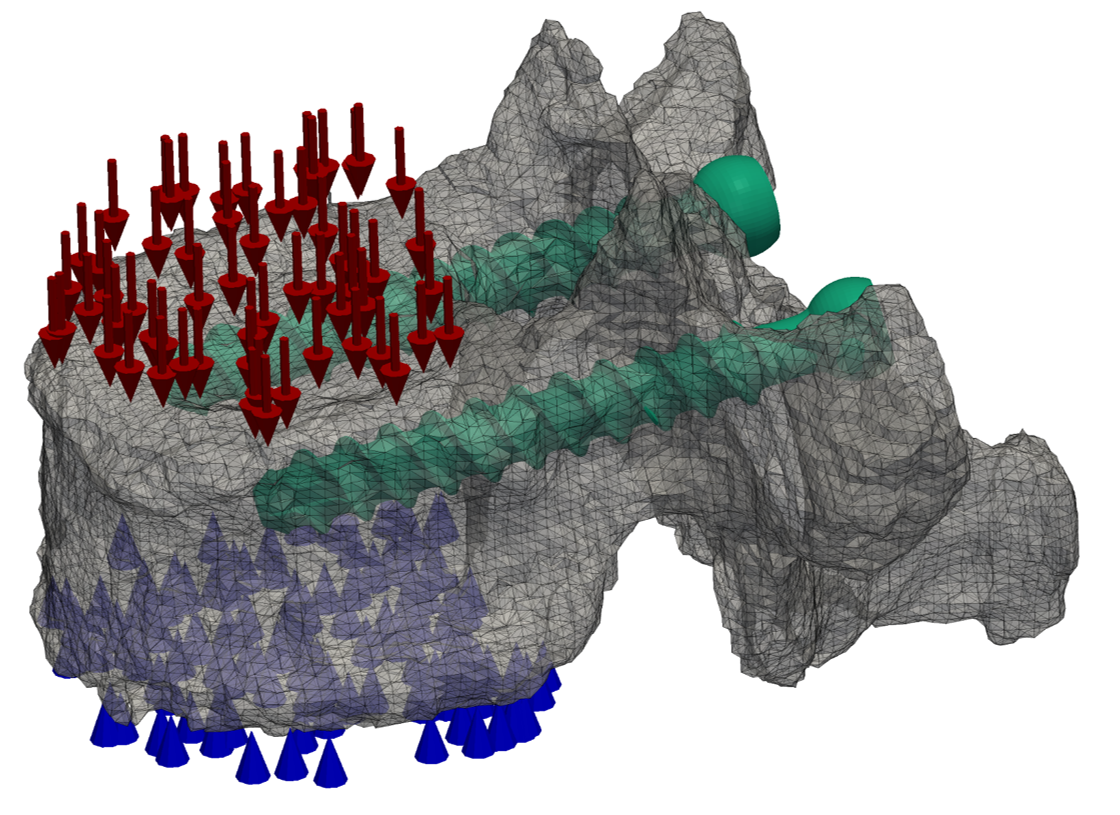}%
        \label{fig::numericalResults::boneLoading}
    }%
    \subfloat[Mesh split for 4 processes.]
        {%
          %\tikzset{external/remake next}%
      \includegraphics[width=0.25\textwidth]%  
        {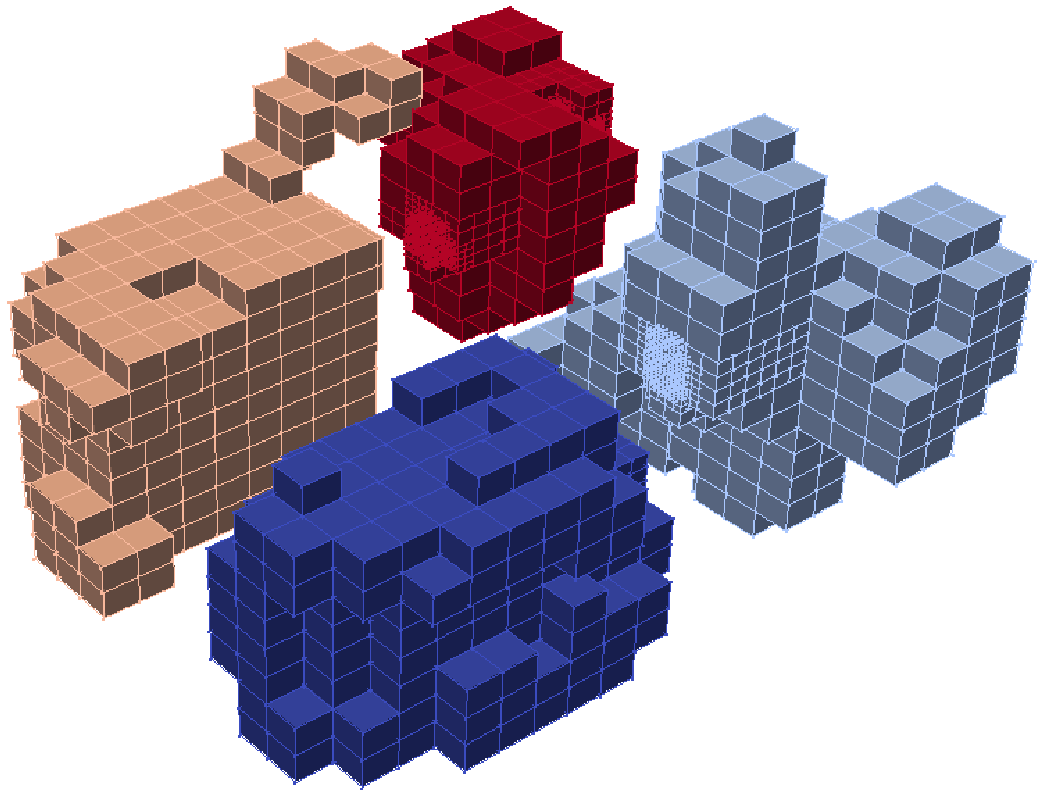}%
          \label{fig::numericalResults::boneMesh}
    }%
    \subfloat[Adaptive refinement.]
    {%
          %\tikzset{external/remake next}%
      \includegraphics[width=0.25\textwidth]{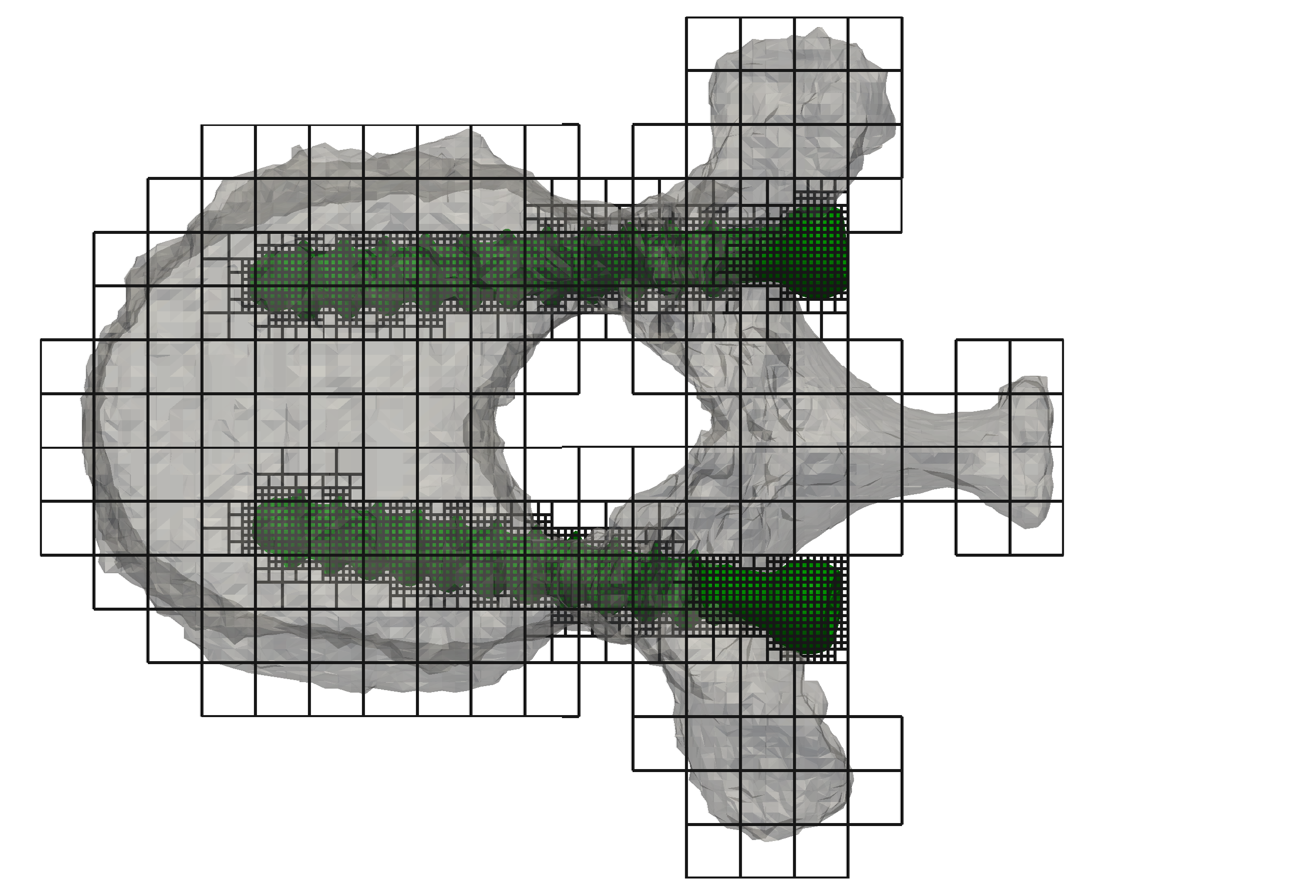}%
      \label{fig::numericalResults::boneRefinement}
    }%
    \subfloat[Composed integration on voxel level.]
    {%
          %\tikzset{external/remake next}%
      \includegraphics[width=0.23\textwidth]{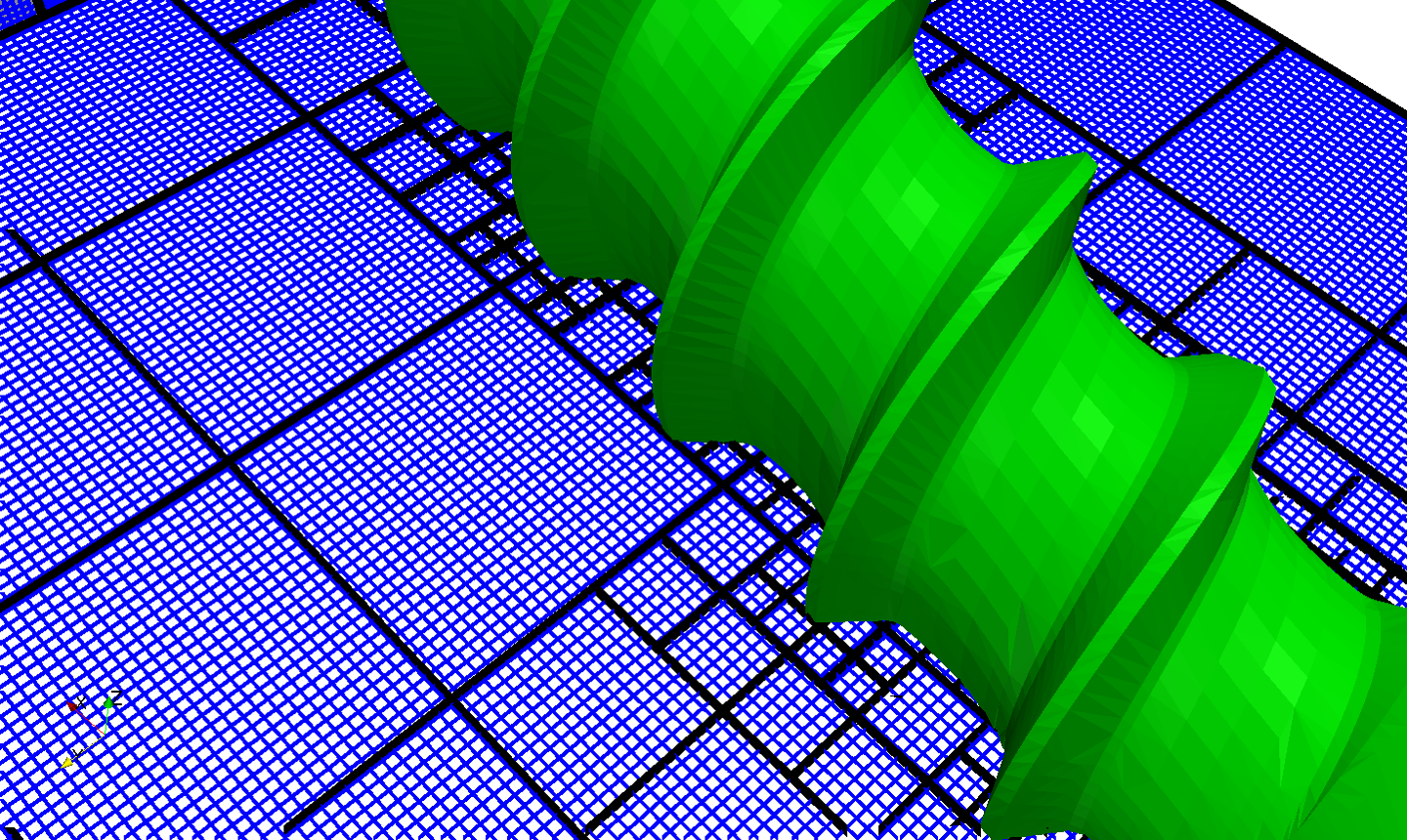}%
      \label{fig::numericalResults::boneIntegration}
    }%
  \caption{Simualtion of a bone-implant system.}
  \label{fig::numericalResults::boneSimulation}
\end{figure}

We conclude this section by considering an example in linear elasticity that combines the finite cell method and multi-level $hp$-refinement to simulate the loading of a bone-implant system. The setup consists of a spinal vertebra with embedded pedicle screws that is supported on its bottom surface and loaded on its upper flank as shown in Figure \ref{fig::numericalResults::boneLoading}. The major benefit of using FCM in this simulation is its ability to deal with multiple geometric models in an easy fashion without the need of generating a boundary conforming mesh. We are thus able to simultaneously compute on the bone material, which is based on a HR-pQCT scan with a voxel size of 146.5 $\mu m$, and the screws, whose geometry is described by a CAD model. A computational mesh shown in Figure \ref{fig::numericalResults::boneMesh} for example, can be easily generated by embedding both the voxel model and screw in a bounding box and using the density values of the CT-scan together with the surface of the screws to filter out elements lying completely in the fictitious domain. The resulting computational domain is refined towards the interface of the bone material and the screws as illustrate in Figure \ref{fig::numericalResults::boneRefinement}, so as to better capture the solution in this area. Furthermore, the composed integration technique illustrated in Figure \ref{fig::multiLevelFCM::integration} is applied on the voxel level, so as to make use of the available fine grain information of the CT-scan. This high resolution integration is depicted in Figure \ref{fig::numericalResults::boneIntegration}, where the black colored cells represent the leaf elements while the blue colored cells the depict the voxels. This results in a very high computational cost, rendering the integration of the linear system the main bottleneck in this simulation.

This example presents a good test for the parallel implementation. The major challenge here lies in the efficient distribution of the computational domain so as to guarantee good performance. The leaf elements not only differ in the number of degrees of freedom due to refinement, but also in the number of voxels. Moreover, an inside-outside test has to be performed for each integration point in order to find out if the point lies within the bone material, screw or fictitious domain. We perform the analysis on an initial mesh comprising 1\,341 base elements that are refined in two steps towards the interface of the bone and the screws. A polynomial degree of $p=3$ is chosen resulting in 7\,571 leaf elements upon refinement and 147\,889 degrees of freedom.      

\begin{figure}[H]
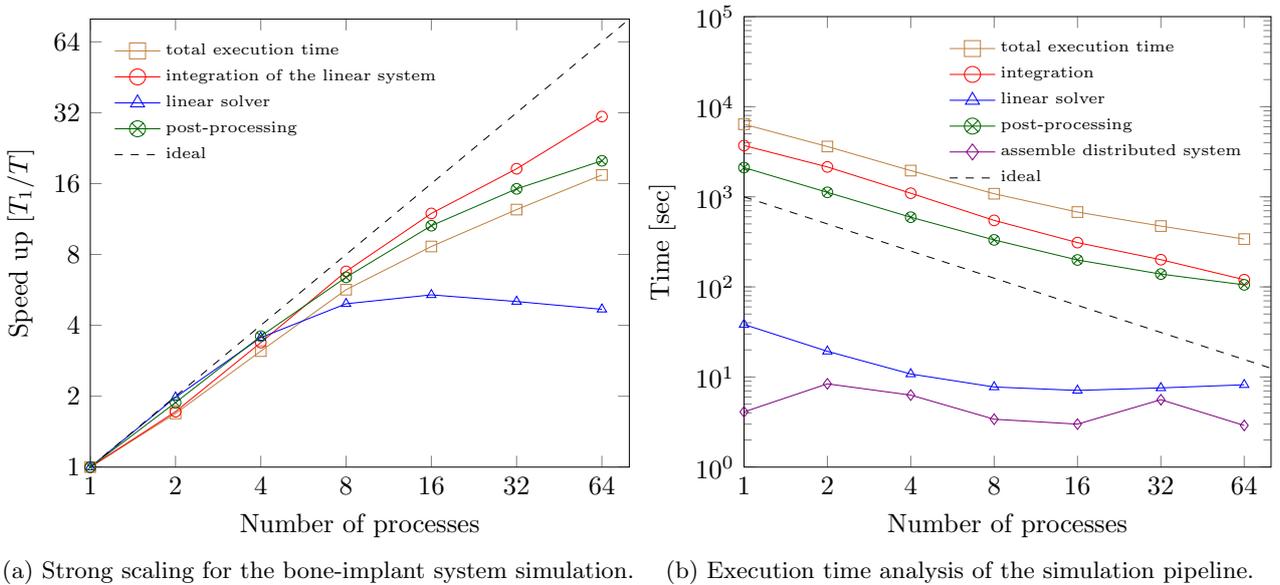

  \begin{center}
   \subfloat[Strong scaling for the bone-implant system simulation.]{%
  %\tikzset{external/remake next}%
    \includetikz[width=\halfTextWidth, height = 7cm]{\graphDir/scalabilityBone}%
    \label{fig::numericalResults::scaleBone}
    }%
   \subfloat[Execution time analysis of the simulation pipeline.]
   {%
   %\tikzset{external/remake next}%
    \includetikz[width=\halfTextWidth, height = 7.3cm]{\graphDir/timingBone}%
    \label{fig::numericalResults::timeBone}
   }%
   \hfill%   
  \end{center} 
  \caption{Analysis of the bone-implant system simulation.}
  \label{fig::numericalResults::strongScalingBone}
\end{figure} 
Figure \ref{fig::numericalResults::strongScalingBone} shows the parallel performance of the different algorithms for the examples at hand. We are able to achieve relatively good scaling results in the most time consuming routines: the integration of the stiffness matrices and the post processing. More effort, however, needs to be invested in load balancing, i.e. in the distribution of the integration domains among processes, in order to improve the algorithms parallel efficiency. 

%For this problem: I also want to show how the parallel implementation preserves
%the exponential convergence of the multi-level hp adaptivity.
%
%\subsection{ Comparison of the different parallel solvers }
%
%\subsubsection{ 3D example eg. Singular Cube Benchmark or the 3D extension of
%the Fichera problem}
%
%I want to show the scalability of pure MPI computations and hybrid computations
%for different problem sizes.
%
%I also want to compare the different parallel solvers in terms of time and
%quality of the solution, MUMPS, SuperLuDist, Hypre and AztecOO solvers.
%
%%\begin{figure}[H]
%%  \begin{center}
%%      %\tikzset{external/remake next}%
%%      \nextfigurename{fcm}%
%%      \includegraphics[width=0.65\textwidth]
%%        {\picsDir/time.png}%
%%  \end{center}
%%  \caption{Performance of different parallel solvers}
%%  \label{fig::parallelImplementation::AdhoC++}
%%\end{figure}
%\subsection{Memory requirements}
%
%I want to show the memory requirements for a duplicated mesh for a mpi-flat and
%hybrid computation. Showing how this limits the number of elements that can be
%used for computations.
%
%\textbf{Graph memory needed over number of mpi-processes}
%
%\subsection{Application in Biomechanical engineering}
%I want to show how the computations performed on the spinal vertebrae, focusing
%on reduced post processing and integration time. We also could compare the two
%post processing strategies employed: All processors write their own file, or all
%processors write into a single file.
%
%\subsection{transient example}

\section{Conclusion and future work}\label{conclusion} 
The article at hand presented a parallelization scheme for the multi-level $hp$-scheme that is adapted to its hierarchical structure. A shared mesh data structure was used in combination with a distributed assembly of the global system, so as to avoid redundant computations of ghost elements. This decision was motivated by the large basis function support of the multi-level $hp$-scheme, which would lead to an unproportionally high number of ghost elements compared to conventional $hp$-schemes. Moreover, the simplicity of a shared mesh data structure allowed for the easy maintenance of mesh consistency among the participating processes. 

Three numerical examples are discussed in this work. They reveal that multi-level $hp$-refinement can be efficiently parallelized and be combined with the finite cell method for the simulation of large engineering problems. Good scalability in the total simulation time for different problem sizes in two and three dimensions is shown. Although each process has to communicate non-local entries of the linear system to other processes during assembly of the global system, this routine does not significantly affect parallel performance and only makes up a small fraction of the total computation time. This influence, however, increases as the number of processes increases due to reduced computational work. The problem size can in this case be increased to counteract this behavior. 

A runtime analysis of the simulation pipeline revealed that the integration of the linear system greatly benefits from the proposed parallel scheme. Perfect scalability is achieved in this algorithm with as little as a few hundred integration domains per process, proving the suitability of the load balancing scheme presented in Algorithm \ref{alg:distribution_of_integration_domains}. This results show the scheme's suitability for problems in which numerical integration dominates. 
Furthermore, the integration scheme shows excellent performance for problems involving both moderate and high polynomial orders.     

The potential of the parallel multi-level $hp$-scheme is demonstrated in the numerical examples considered. Further numerical studies are, however, necessary to extend the schemes applicability and improve its performance. Although the memory requirements per process significantly decrease with an increase in the number of processes, this behavior does not hold when high numbers of MPI processes are used, but levels off when the size of the computational mesh begins to dominate. The maximum size of the computation mesh that can be replicated on all processes is, therefore, limited. The onset of this leveling-off can be delayed by optimizing the memory footprint for the code, thus increasing the maximum size of the duplicated mesh. A more feasible solution to this problem would be the use of a distributed mesh structure, as suggested in \cite{bangerth_2008}. Moreover, utilizing the scheme in a hybrid framework would help reduce the amount of memory needed on a single node. The scalability of the parallel scheme is greatly governed by the linear solver as shown in the numerical examples. A more in depth analysis has to be conducted to find the best solver configurations for the problem types considered. The final numerical example showed the advantage of using the multi-level $hp$-scheme in conjunction with the finite cell method. The efficiency of the load balancing algorithm has to be improved in this case and adapted to the voxel based integration scheme. Yet another research direction would be the extension of the parallel scheme's application field. One possibility would be the use of this scheme in nonlinear problems which require complex three dimensional refinement patterns.

%\input{materialInterface1D/materialInterface1D}
%\input{example/example}

%% Acknowledgements -------------------------------  
\section*{Acknowledgements} 
The first and the last author gratefully acknowledge the
financial support of the German Research Foundation (DFG) under
Grant RA 624/27-1.

%\newpage 
%% References -------------------------------------
%\bibliographystyle{apalike}
\bibliographystyle{ieeetr}
\bibliography{paper}

\end{document}